\documentclass[manuscript,screen]{acmart}
\AtBeginDocument{%
  }

\setcopyright{acmlicensed}
\copyrightyear{2025}
\acmYear{2025}
\acmDOI{XXXXXXX.XXXXXXX}
\acmISBN{978-1-4503-XXXX-X/2018/06}

\usepackage{amsthm}
\usepackage{bm}
\usepackage[linesnumbered,ruled,vlined]{algorithm2e}
\usepackage{graphicx}
\usepackage{amsmath}

\usepackage{multirow}
\usepackage{makecell}
\usepackage{subcaption}
\usepackage{caption}
\usepackage{listings}
\usepackage{multicol}
\usepackage{wrapfig}
\usepackage{enumitem}

\begin{document}

\title{Efficient GPU-Centered Singular Value Decomposition Using the Divide-and-Conquer Method}

\author{Shifang Liu}
\email{liushifang@iscas.ac.cn}
\orcid{https://orcid.org/0009-0003-9037-4902}
\authornotemark[1]
\author{Huiyuan Li}
\email{huiyuan@iscas.ac.cn}
\orcid{https://orcid.org/0000-0002-6326-9926}
\author{Hongjiao Sheng}
\email{hongjiao@iscas.ac.cn}
\orcid{https://orcid.org/0009-0005-7424-8580}
\affiliation{%
  \institution{State Key Laboratory of Computer Science/Laboratory of Parallel Software and Computational Science, Institute of Software Chinese Academy of Sciences}
  \city{Beijing}
  \country{China}}
  
\author{Haoyuan Gui}
\email{guihaoyuan123@icloud.com}
\orcid{https://orcid.org/0000-0002-1218-7920}
\author{Xiaoyu Zhang}
\email{zhangxy420@foxmail.com}
\orcid{https://orcid.org/0009-0008-3477-7359}
\affiliation{%
  \institution{Institute of Software Chinese Academy of Sciences and University of Chinese Academy of Sciences}
  \city{Beijing}
  \country{China}}

\renewcommand{\shortauthors}{S. Liu et al.}

\begin{abstract}
Singular Value Decomposition (SVD) is a fundamental matrix factorization technique in linear algebra, widely applied in numerous matrix-related problems. However, traditional SVD approaches are hindered by slow panel factorization and frequent CPU-GPU data transfers in heterogeneous systems, despite advancements in GPU computational capabilities. 
In this paper, we introduce a 
GPU-centered SVD algorithm, incorporating a novel GPU-based  bidiagonal divide-and-conquer (BDC) method. 
We reformulate the algorithm and data layout of different steps for SVD computation, performing all panel-level computations and trailing matrix updates entirely on GPU to eliminate CPU-GPU data transfers. 
Furthermore, we integrate related computations to optimize BLAS utilization, thereby increasing arithmetic intensity and fully leveraging the computational capabilities of GPUs. 
Additionally, we introduce a newly developed GPU-based BDC algorithm that restructures the workflow to eliminate matrix-level CPU-GPU data transfers and enable asynchronous execution between the CPU and GPU.
Experimental results on AMD MI210 and NVIDIA V100 GPUs demonstrate that our proposed method achieves speedups of up to 1293.64x/7.47x and 14.10x/12.38x compared to rocSOLVER/cuSOLVER and MAGMA, respectively.

\end{abstract}




\maketitle

\section{Introduction}

Singular Value Decomposition (SVD) is a fundamental operation in linear algebra, widely used for computing the pseudoinverse of a matrix, solving homogeneous linear equations, addressing total least squares minimization problems, and finding approximation matrices. 
It has been successfully applied to various fields, such as bioinformatics \citep{troyanskaya2001missing,wall2003singular}, physics \citep{hoecker1996svd,frankenberg2014prospects}, and machine learning \citep{zhang2010discriminative,candes2011robust,liu2011robust}. 
In particular, the SVD of tall-and-skinny (TS) matrices—where the number of rows significantly exceeds the number of columns—has attracted considerable attention in various fields, including computer vision \citep{oh2015fast}, image compression \citep{andrews1976singular,sadek2012svd}, facial recognition \citep{turk1991face,zhang2005new,zhang2010discriminative}, and data analysis \citep{henry19928,ding2010application,harkat2006improved}.
Precisely, the SVD of an $m \times n$ matrix $\bm{A}$ is given by
\begin{equation}
    \bm{A} = \bm{U} \bm{\Sigma} \bm{V}^\mathrm{T}, \ \text{with} \ m \geq n,
\end{equation}
where $\bm{\Sigma} = \texttt{diag}(\sigma_1, \sigma_2, \dots, \sigma_n)$ is an $m \times n$ diagonal matrix with real, non-negative entries $\sigma_1 \geq \sigma_2 \geq \cdots \geq \sigma_n \geq 0$, representing the singular values of $\bm{A}$. 
$\bm{U}$ and $\bm{V}$ are $m \times m$ and $n \times n$ orthogonal matrices, representing the left and right singular vectors of $\bm{A}$, respectively.

With the increasing demand for high-performance computing (HPC), optimizing SVD on GPU has gained significant attention \citep{lahabar2009singular,gates2018accelerating,tomov2010accelerating,ltaief2009parallel}.
AMD’s rocSOLVER \citep{rocsolver} provides an initial SVD solver as part of the Radeon Open Compute platform (ROCm) \citep{rocm}, offering only an interface for computing the SVD of a bidiagonal matrix using the QR method. 
Similarly, NVIDIA’s cuSOLVER \citep{cusolver} also offers an interface based on the QR method.
However, neither library currently offers an interface for the more efficient bidiagonal divide-and-conquer (BDC) method.
Moreover, the MAGMA \citep{magmahome} library provides interfaces for both the QR and BDC methods. It supports CPU+GPU heterogeneous compute nodes, leveraging the strengths of both processing units to optimize performance for SVD and other linear algebra operations.

In heterogeneous compute nodes, 
CPUs excel at low-latency, sequential tasks through deep memory hierarchies and instruction-level parallelism, while GPUs deliver high throughput for data- and thread-parallel operations. 
Modern matrix factorization algorithms typically split each iteration into two steps:
(1) panel factorization, which is relatively slow but involves small matrices, and (2) trailing matrix update, which is fast and involves large matrix operations.
In the MAGMA framework \citep{dongarra2014matrix}, panel factorizations are executed on the CPU, while trailing matrix updates are offloaded to the GPU.
Due to the algorithmic pipeline, trailing matrix updates are typically fully or partially overlapped with panel factorizations and CPU-GPU data transfers. However, they are not the primary performance bottleneck.
Meanwhile, with advancements in GPU computational power, the imbalance between the speed of computations and CPU-GPU data transfers has been further exacerbated, such that even high computation-intensive kernels can be dominated by the costs associated with data transfers. 
As a result, even compute-intensive kernels may be dominated by data movement overheads, limiting overall  efficiency in workloads with frequent CPU-GPU interactions.


To tackle the bottlenecks, 
we propose a GPU-centered SVD algorithm that restructures computation and data layout to maximize GPU efficiency. 
Specifically, We introduce a merged-rank-(2$b$) bidiagonalization strategy that performs both panel factorization and trailing matrix updates entirely on GPU, eliminating CPU-GPU data transfers.
Furthermore, by merging computations, this approach increases arithmetic intensity, thereby improving GPU utilization.
For the other stages, including QR factorization and back-transformations, panel factorization is also performed on GPU, utilizing a modified CWY transform \citep{puglisi1992modification} to enhance compute-bound BLAS3 operations, substantially increasing arithmetic intensity and fully exploiting GPU computational capabilities. 
Furthermore, we propose a new GPU-based BDC algorithm that eliminates matrix-level data transfers and enables asynchronous CPU-GPU execution for further acceleration.
We consider the main contributions of this paper to be:
\begin{itemize}
     \item 
    We reformulate the algorithm and data layout for the SVD computation steps—bidiagonalization, QR factorization, and back-transformations—executing all panel-level computations and trailing matrix updates entirely on GPU to eliminate CPU-GPU data transfers. Additionally, we integrate related computations to optimize BLAS utilization, 
    maximizing the exploitation of GPU computational capabilities.

    \item We introduce a new efficient GPU-based BDC algorithm that eliminates matrix-level data transfers and enables asynchronous execution between CPU and GPU.
    
    \item We conduct extensive experiments across various matrix sizes and GPUs to demonstrate the efficiency of the proposed SVD algorithm.
    Compared to cuSOLVER/rocSOLVER and MAGMA, the speedup reaches up to 1293.64x/7.47x and 14.10x/12.38x on AMD MI210 and NVIDIA V100 GPUs, respectively.
\end{itemize}

The rest of the paper is organized as follows: Section 2 reviews related work. 
Section 3 outlines the experiment setup. 
Section 4 provides details on our method and optimization strategies, along with some related experiment results.
Section 5 evaluates our implementations and shows end-to-end SVD performance. 
Section 6 concludes this paper.

\section{Related work}
Theoretically, the singular values are the square roots of the eigenvalues of $\bm{A}^\mathrm{T}\bm{A}$, the columns of $\bm{V}$ are the eigenvectors of $\bm{A}^\mathrm{T}\bm{A}$, and the columns of $\bm{U}$ are the eigenvectors of $\bm{A}\bm{A}^\mathrm{T}$. However, this approach is not ideal for computation, as roundoff errors in the calculation of $\bm{A}^\mathrm{T}\bm{A}$ and $\bm{A}\bm{A}^\mathrm{T}$ frequently result in the loss of relevant information. 
There are two dominant categories of SVD algorithms for dense matrices: Jacobi methods and bidiagonalization methods.

Jacobi methods apply plane rotations to the entire matrix $\bm{A}$. Two-sided Jacobi methods, first proposed by Kogbetliantz in 1955 \citep{kogbetliantz1955solution}, iteratively apply rotations on both sides of $\bm{A}$ to bring it to diagonal form, while one-sided Jacobi methods, proposed by Hestenes in 1958 \citep{hestenes1958inversion}, apply rotations on one side to orthogonalize the columns of $\bm{A}$, implicitly bringing  $\bm{A}^\mathrm{T} \bm{A}$ to diagonal. Although Jacobi methods are generally slower than bidiagonalization methods, they remain of interest due to their simplicity, ease of parallelization, and potentially better accuracy for certain classes of matrices.

Golub and Kahan in 1965  \citep{golub1965calculating}  proposed the first stable SVD algorithm for computers using a bidiagonlization method. 
Golub and Reinsch  \citep{golub1971singular} realized the first implementation in Algol60, the programming language of the time.
The classical bidiagonalization method proceeds in the following three stages:
\begin{enumerate}
    \item Bidiagonal reduction: Reduce $\bm{A} \in \mathbb{R}^{m \times n}$ to a bidiagonal form $\bm{A} = \bm{U}_1 \bm{B} \bm{V}_1^\mathrm{T}$ by applying a series of 
    orthogonal similarity 
    transformations, where $\bm{U}_1$ and $\bm{V}_1$ are orthogonal matrices, and $\bm{B}$ is a real upper bidiagonal matrix when $m \geq n$. 
    \item Diagonalization: Compute the bidiagonal SVD as $\bm{B}=\bm{U}_2\bm{\Sigma}\bm{V}_2^\mathrm{T}$, where $\bm{U}_2$ and $\bm{V}_2$ are orthogonal matrices, and $\bm{\Sigma}$ is a diagonal matrix.
    \item Singular vector back-transformation:
     The singular vectors of $\bm{A}$ can be computed as $\bm{U}=\bm{U}_1 \bm{U}_2$ and $\bm{V}^\mathrm{T}=\bm{V}_2^\mathrm{T} \bm{V}_1^\mathrm{T}$.
\end{enumerate}

The bidiagonal reduction is the most compute-intensive step  in SVD, requiring approximately $O\left(\frac{8}{3}n^3\right)$ floating-point operations, and can be performed using either a one-stage or two-stage approach.
In the one-stage method, $\bm{A}$ is decomposed as $\bm{A} = \bm{U}_1 \bm{B} \bm{V}_1^\mathrm{T}$ by applying a series of Householder transformations, where $\bm{B}$ is bidiagonal matrix, and $\bm{U}_1$ and $\bm{V}_1$ are orthogonal matrices. 
An early GPU-accelerated implementation of one-stage bidiagonal reduction followed by QR-based bidiagonal SVD was proposed by Lahabar and Narayanan \citep{lahabar2009singular}. 
The current GPU-accelerated one-stage implementation in MAGMA was introduced by Tomov et al. \citep{tomov2010accelerating}.
However, the one-stage reduction relies heavily on memory-bound BLAS2 operations. 
To mitigate this, Gr{\"o}sser and Lang \citep{grosser1999efficient} proposed a two-stage reduction:
first reducing $\bm{A}$ to a band matrix, $\bm{A} = \bm{U}_a \bm{\hat{A}} \bm{V}_a^\mathrm{T}$, followed by a second reduction to bidiagonal form, $\bm{\hat{A}} = \bm{U}_b \bm{B} \bm{V}_b^\mathrm{T}$ \citep{lang1996parallel}. 
Although it involves more operations than the one-stage algorithm, the first stage leverages efficient BLAS3 operations, making it efficient than the one-stage bidiagonal reduction.
Ltaief et al. implemented the first \citep{ltaief2009parallel} and second stages \citep{ltaief2013high} using tile algorithms with dynamic scheduling for multi-core CPUs in PLASMA \citep{plasmahome},
with later optimizations by
Haidar et al. \citep{haidar2013improved,haidar2012comprehensive}.
Gates et al. \citep{gates2018accelerating} further accelerated the first stage with a GPU while employed the PLASMA CPU implementation for the second stage.
Two-stage reduction also requires using two corresponding singular vector back-transformation steps, first multiplying $\bm{U}_b \bm{U}_2$ and $\bm{V}^\mathrm{T}_2 \bm{V}^\mathrm{T}_b$, then multiplying $\bm{U}_a ( \bm{U}_b \bm{U}_2 )$ and $(\bm{V}^\mathrm{T}_2 \bm{V}^\mathrm{T}_b ) \bm{V}^\mathrm{T}_a$, when the singular vectors are required. 
A further drawback of the two-stage reduction is that the orthogonal transformations used in the band-to-bidiagonal process must be accumulated into an orthogonal matrix, which can be challenging to perform efficiently due to the irregular nature and fine granularity of the operations introduced in the second stage.
Given these complexities, we choose the one-stage bidiagonal reduction algorithm in our method.

After the bidiagonal reduction, several algorithms exist for computing the bidiagonal SVD. The original method is QR iteration \citep{lahabar2009singular,demmel1990accurate,golub2013matrix}. Later developments include BDC \citep{gu1995divide} and multiple relatively robust representations (MRRR) \citep{willems2006computing}.
The BDC algorithm enhances performance in two key ways: it reduces the complexity of bidiagonal SVD to $\frac{8}{3}n^3$, potentially achieving $O(n^{2.3})$ or lower \citep{tisseur1999parallel}, and it replaces the memory-bandwidth-limited BLAS2 Givens rotations of QR iteration, which require approximately $12n^3$ operations, with more efficient BLAS3 operations.
MRRR further improves efficiency by lowering the complexity of the bidiagonal SVD to $O(n^2)$. However, a stable MRRR version for the SVD is not yet available in libraries like LAPACK. Therefore, we chose to examine BDC in our method. 

For TS matrix ($m\gg n$), it is more efficient to first perform a QR factorization of $\bm{A}$ and then compute the SVD of the $n \times n$ matrix $\bm{R}$, since if $\bm{A}=\bm{Q}\bm{R}$ and $\bm{R}=\bm{U}_0\bm{\Sigma}\bm{V}_0^\mathrm{T}$, then the SVD of $\bm{A}$ is given by $\bm{A} =(\bm{Q}\bm{U}_0)\bm{\Sigma}\bm{V}_0^\mathrm{T}$. Chan \citep{chan1982improved} analyzed this optimization, showing that it reduces the number of floating-point operations.
The most widely used approach for QR factorization is based on Householder transformations. 
To enable efficient implementation using high-performance matrix-matrix operations, two formulations have been proposed for accumulating multiple Householder reflectors: the WY transform \citep{bischof1987wy} and the CWY transform \citep{schreiber1989storage}. In addition, several modifications to the CWY transform \citep{walker1988implementation,puglisi1992modification,joffrain2006accumulating} have been introduced to improve its performance.
In this paper, our method utilizes the modified CWY transform, 
further optimized for GPU architectures to maximize the exploitation of GPU computational capabilities.

\begin{figure}[ht]
 \centering
    \includegraphics[width=0.62\linewidth]{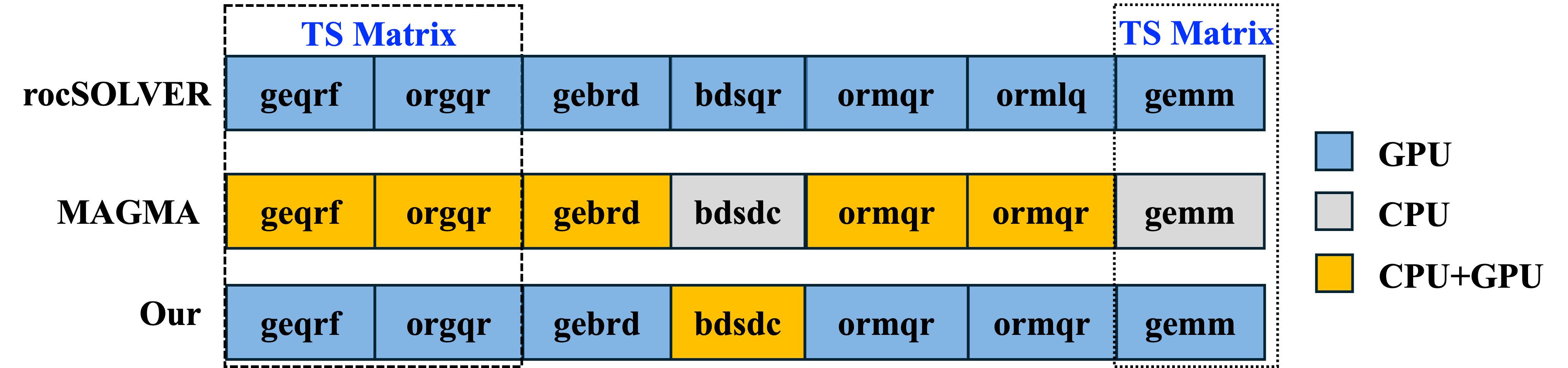}
    \caption{Execution profile of SVD between rocSOLVER, MAGMA and our method.}
    \label{fig:svd-magma-our-execution}
\end{figure}

In this paper, we accelerate all phases of the SVD algorithm on GPU. 
Fig.~\ref{fig:svd-magma-our-execution} presents the execution profile of the overall SVD solver for rocSOLVER, MAGMA, and our method, with phases named in a manner consistent with LAPACK routines. 
As shown, the rocSOLVER implementation executes all phases entirely on GPU but utilizes QR iteration (\texttt{bdcqr}) for the diagonalization phase, as \texttt{bdcdc} has not been implemented. 
Our SVD method also executes all phases on GPU, except for the \texttt{bdcdc} phase, which employs a CPU+GPU heterogeneous approach without matrix-level data transfers. 
In contrast, MAGMA primarily relies on a CPU+GPU heterogeneous execution model across most phases, with both \texttt{bdcdc} and final back-transformation of singular vectors (\texttt{gemm}) for TS matrices executed on CPU.

\section{Experiment Setup}
Our experiments are conducted on a Linux system (version 3.10.0-1062.4.1.el7.x86\_64) with an Intel Xeon Gold 6154 CPU.
We evaluate performance on two accelerators: AMD Instinct MI210 and NVIDIA Tesla V100-PCIe.
The MI210, based on the 6 nm CDNA2 architecture, features 64 GB of HBM2e memory with 1.6 TB/s bandwidth 
and delivers up to 22.6 TFLOPS of peak performance in FP64/FP32.
It operates under ROCm 5.7.0 (driver version 5.16.9.22.20), which provides a C++ compiler and optimized libraries such as rocBLAS and rocSOLVER. 
The V100, based on the 12 nm Volta architecture, provides up to 32 GB of HBM2 memory  
with 900 GB/s bandwidth
and delivers peak performance of 7.8 TFLOPS in FP64 and 15.7 TFLOPS in FP32.
It is supported by CUDA 12.0 (driver version 525.60.13), along with the cuBLAS and cuSOLVER libraries. 
For comparison, we benchmark our algorithm against the state-of-the-art MAGMA library (version 2.8.0) on both accelerators. All tests are performed in double precision and utilize all 10 CPU cores.
For the SVD experiments, we use MAGMA’s matrix generation routine ($\texttt{magma\_generate\_matrix}$) to create random matrices with specified condition numbers and singular value distributions. We consider four matrix types:
\begin{itemize}
\item \texttt{random}: matrix entries are uniformly distributed in the range $(0, 1)$, serving as the default test case in this paper.
\item  \texttt{SVD\_logrand}($\theta$):  singular values $\log(\sigma_i)$ are uniformly distributed over $(\log(\frac{1}{\theta}), \log(1))$.
\item  \texttt{SVD\_arith}($\theta$): singular values are arithmetically distributed as $\sigma_i = 1 - \frac{(i - 1)}{(n - 1)}(1 - \frac{1}{\theta})$.
\item  \texttt{SVD\_geo}($\theta$): singular values are geometrically distributed as $\sigma_i = \theta^{-\frac{(i-1)}{(n-1)}}$.
\end{itemize}
Here, the notation \texttt{SVD\_‘NAME’} indicates that the singular values of the generated matrix follow the specified ‘NAME’ distribution, and $\theta$ denotes the condition number.

\section{SVD Algorithm}\label{sec:svd-overall}

\subsection{Bidiagonalization}
\subsubsection{\textbf{Algorithm}}
For a nonzero vector $\bm{v} = (v_1, v_2, \dots, v_n)^{\rm{T}} \in \mathbb{R}^n$, a Householder reflector is defined as $\bm{H} = \bm{I} - \tau \bm{y}\bm{y}^{\rm{T}}$, where $\bm{I}$ is the identity matrix, $\tau = \frac{2}{\|\bm{v}\|_2^2}$ is a scalar, and the Householder vector is given by $\bm{y} = (v_1 \pm \|\bm{v}\|_2, v_2, \dots, v_n)^{\rm{T}}$.
Applying $\bm{H}$ to $\bm{v}$ yields $\bm{H}\bm{v}=-\texttt{sign}(v_1)\|\bm{v}\|_2\bm{e}_1$,
where $\bm{e}_1$ is the first standard basis vector.
In the bidiagonalization step, two orthogonal matrices, $\bm{U}_1$ and $\bm{V}_1$, are applied to the left and right of $\bm{A} \in \mathbb{R}^{m \times n}$ to reduce it to bidiagonal form: $\bm{B} = \bm{U}_1^\mathrm{T} \bm{A} \bm{V}_1$.  
The matrices $\bm{U}_1$ and $\bm{V}_1$ are represented as products of elementary Householder reflectors
\begin{equation}\label{eq:gebrd-house}
 {\bm{U}_1 = {\textstyle \prod_{i=1}^{n}}\bm{H}_i\ \text{and} \ 
    \bm{V}_1 = {\textstyle \prod_{i=1}^{n-1}}\bm{G}_i.}
\end{equation}
Each $\bm{H}_i$ and $\bm{G}_i$ has the form: $ \bm{H}_i = \bm{I} -  \tau_i \bm{v}_i \bm{v}^\mathrm{T}_i \ \text{and} \ \bm{G}_i = \bm{I} - \pi_i \bm{u}_i \bm{u}^\mathrm{T}_i $,
where $\tau_i$ and $\pi_i$ are scalars, and $\bm{v}_i$ and $\bm{u}_i$ are Householder vectors. 
$\bm{H}_i$ eliminates elements below the diagonal in column $i$, while $\bm{G}_i$ eliminates elements right of the off-diagonal in row $i$. 
Then, update the trailing matrix after every column-row elimination.
Let $\bm{A}_{(i-1)}$ be the reduced matrix after step $i-1$. Applying $\bm{H}_i$ and $\bm{G}_i$ on the left and right yields
\begin{equation}\label{eq:tail-nonblock}
    \bm{A}_{(i)}=\bm{H}_i \bm{A}_{(i-1)} \bm{G}_i = \bm{A}_{(i-1)} - \bm{v}_i \bm{y}_i^\mathrm{T} - \bm{x}_i \bm{u}_i^\mathrm{T},
\end{equation}
where $\bm{y}_i = \tau_i \bm{A}^\mathrm{T}_{(i-1)} \bm{v}_i \ \text{and} \
\bm{x}_i =  \pi_i \left( \bm{A}_{(i-1)} - \bm{v}_i \bm{y}^\mathrm{T}_i \right) \bm{u}_i$.

\begin{figure}[ht] 
    \centering
   \includegraphics[width=0.55\linewidth]{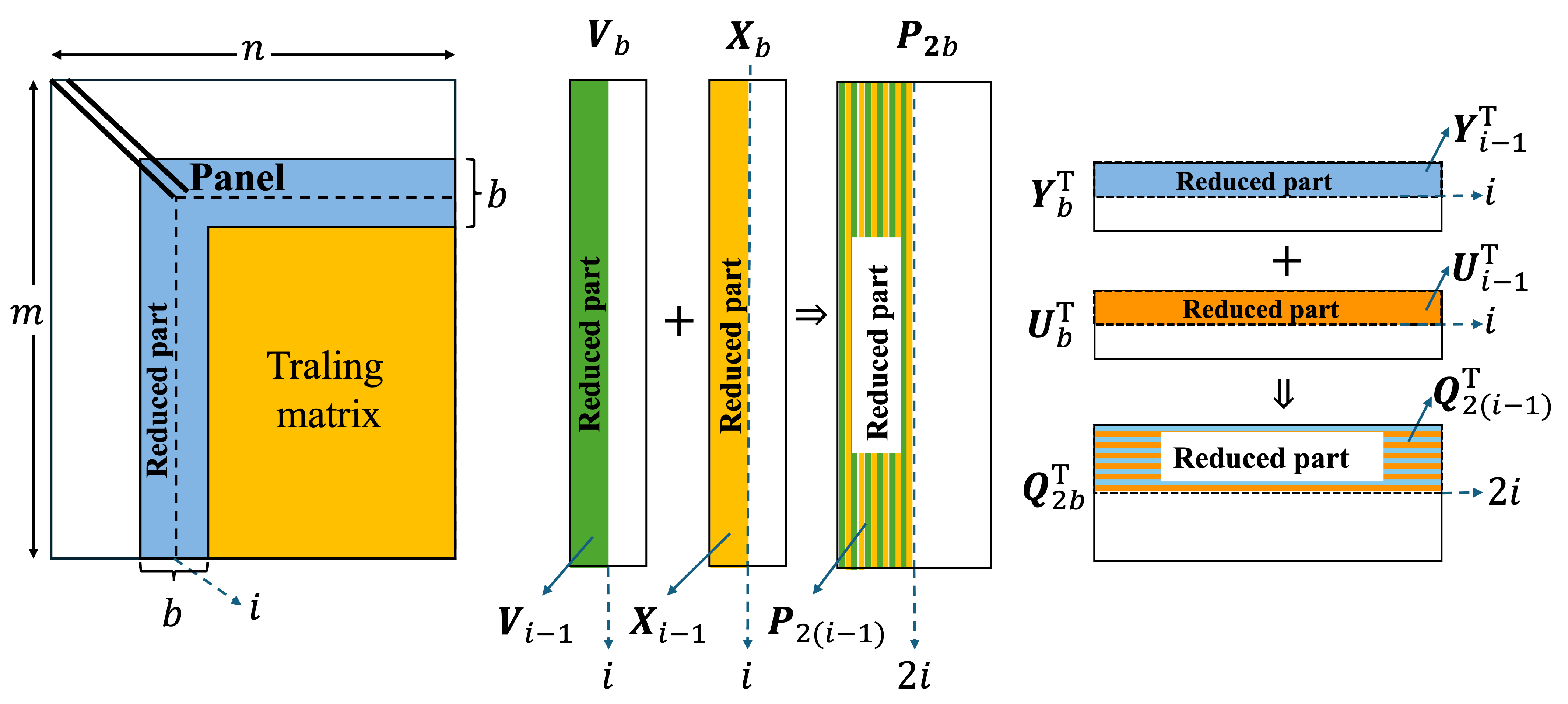}
  \caption{Bidiagonalization blocked algorithm.}
  \label{fig:bi-diag-block}  
\end{figure}

The transformation in \eqref{eq:tail-nonblock} is a rank-2 update that involves memory-bandwidth-limited BLAS2 operations.
To address this, the trailing matrix update can be deferred by first performing bidiagonalization on a block of columns and rows, followed by a delayed update of the trailing matrix using the WY representation \cite{bischof1987wy}, as illustrated on the left side of Fig.~\ref{fig:bi-diag-block} and implemented in the LAPACK routine \texttt{gebrd}. 
Blocking together $b$ reflectors of \eqref{eq:tail-nonblock}, we obtain:
 \begin{equation}
 \label{eq:bidiag-A-trail}
   \bm{A}_{(i)}=\bm{H}_b \cdots \bm{H}_1\bm{A} \bm{G}_1 \cdots \bm{G}_b 
   = \bm{A} - 
   \bm{V}_b \bm{Y}_b^\mathrm{T} - 
   \bm{X}_b \bm{U}_b^\mathrm{T},
 \end{equation}
where $\bm{V}_b=[\bm{v}_1, \cdots,\bm{v}_b]$, and similarly with $\bm{Y}_b$, $\bm{X}_b$ and $\bm{U}_b$.
Evidently, which needs two matrix-matrix multiplications (\textbf{\texttt{gemm}}$\times$\textbf{2}) to update the trailing matrix (called rank-$2b$ update).
Note that it is possible to update just part of $\bm{A}$ within the panel, namely, the $i$-th column and $i$-th row of $\bm{A}$, in order to process with the computation of the $\bm{H}_i$ and $\bm{G}_i$. Hence, a delayed update becomes possible.
Consequently, the computation of vector $\bm{y}_i$ needs to be changed to
\begin{equation}
\label{eq:bidiag-y-k}
 \bm{y}_i = \tau_i \bm{A}^\mathrm{T}_{(i-1)} \bm{v}_i 
 = \tau_i \left( \bm{A} - \bm{V}_{i-1}\bm{Y}_{i-1}^\mathrm{T} -\bm{X}_{i-1} \bm{U}_{i-1}^\mathrm{T} \right)^\mathrm{T} \bm{v}_i 
 = \tau_i  \bm{A}^\mathrm{T} \bm{v}_i - 
 \tau_i \bm{Y}_{i-1} \left(\bm{V}_{i-1}^\mathrm{T} \bm{v}_i \right)  -
 \tau_i \bm{U}_{i-1} \left(\bm{X}_{i-1}^\mathrm{T} \bm{v}_i\right).  
\end{equation}
Obviously, each iteration involves one matrix-vector product (\texttt{gemv}) with the full trailing matrix and four tall-and-skinny matrix-vector products (\textbf{\texttt{gemv}}$\times$\textbf{4}) to compute $\bm{y}_i$.
The computation of $\bm{x}_i$ needs to change similarly,
\begin{equation}\label{eq:bidiag-x-k}
  \bm{x}_i = \pi_i \left( \bm{A}_{(i-1)} - \bm{v}_i \bm{y}^\mathrm{T}_i \right) \bm{u}_i 
  = \pi_i \left( \bm{A} - \bm{V}_{i} \bm{Y}_{i}^\mathrm{T}-\bm{X}_{i-1} \bm{U}_{i-1}^\mathrm{T} \right) \bm{u}_i
   = \pi_i  \bm{A} \bm{u}_i - 
   \pi_i \bm{V}_{i} \left(\bm{Y}_{i}^\mathrm{T}\bm{u}_i \right)-
   \pi_i \bm{X}_{i-1}  \left(\bm{U}_{i-1}^\mathrm{T} \bm{u}_i \right) .
\end{equation}

Further, we can find that \eqref{eq:bidiag-y-k} can be merged to
\begin{equation}
\label{eq:bidiag-y-k2}
 \bm{y}_i =\tau_i \bm{A}^\mathrm{T} \bm{v}_i - \tau_i \left( 
     \left[ \begin{matrix}
    \bm{Y}_{i-1}, \bm{U}_{i-1}
\end{matrix} \right] \left[ \begin{matrix}
    \bm{V}_{i-1}, \bm{X}_{i-1}
\end{matrix} \right]
^\mathrm{T}
\right) \bm{v}_i.   
\end{equation} 
Let $\bm{P}_{2b}= [\bm{v}_1, \bm{x}_1, \bm{v}_2, \bm{x}_2, \cdots, \bm{v}_b, \bm{x}_b]$ and $\bm{Q}_{2b} = [\bm{y}_1, \bm{u}_1, \bm{y}_2, \bm{u}_2, \cdots, \bm{y}_b, \bm{u}_b]$. 
If $\bm{P}_{2(i-1)}$ and $\bm{Q}_{2(i-1)}$ are the reduced parts after step $(i-1)$, as shown in Fig.~\ref{fig:bi-diag-block}, then \eqref{eq:bidiag-y-k2} can be restructured to
\begin{equation}\label{eq:bidiag-y-k3}
    \bm{y}_i
    = \tau_i \bm{A}^\mathrm{T} \bm{v}_i - 
    \tau_i {\bm{Q}_{2(i-1)}} \left(\bm{P}_{2(i-1)}^\mathrm{T} \bm{v}_i \right),
\end{equation}
which combines the four TS matrix-vector products (\textbf{\texttt{gemv}}$\times$\textbf{4}) in each iteration into two matrix-vector products (\textbf{\texttt{gemv}}$\times$\textbf{2}). 
Similarly, \eqref{eq:bidiag-x-k} can be combined into
\begin{equation}
\label{eq:bidiag-x-k3}
 \bm{x}_i
    =\pi_i \bm{A}\bm{u}_i - 
    \pi_i \bm{P}_{2i-1}  \left({\bm{Q}^\mathrm{T}_{2i-1}} \bm{u}_i\right).
\end{equation}

Furthermore, the trailing matrix update in \eqref{eq:bidiag-A-trail} can be rearranged as follows:
\begin{equation}\label{eq:bidiag-A-trail3}
     \bm{A}_{(i)} = \bm{A} - \bm{P}_{2b} \bm{Q}_{2b}^\mathrm{T},
\end{equation}
which merges two matrix-matrix multiplications (\textbf{\texttt{gemm}}$\times$\textbf{2}) into one (\textbf{\texttt{gemm}}$\times$\textbf{1}) to update the trailing matrix (called merged-rank-$(2b)$ update).
\begin{algorithm}[ht]\small
\caption{A pseudocode of our proposed merged-rank-$(2b)$ \texttt{gebrd} algorithm}\label{alg:bi-diag-block}

\textbf{function} \textcolor{blue}{\texttt{gebrd}}$(A)$\\
    \For{$i=1:n:b $}{
        (1) $(P, Q)=\textcolor{blue}{\texttt{labrd}}\left(A_{i:m,i:n} \right)$; \CommentSty{//reduce row and column panel to bidiagonal form}\\
        
        (2) $A_{i+b:m,i+b:n} {=} A_{i+b:m,i+b:n} - {P}  {Q}^\mathrm{T}$; \CommentSty{//{update the tailing matrix (\textbf{gemm}$\times$\textbf{1}})}
    }
\textbf{end function}\\
\textbf{function} $\textcolor{blue}{\texttt{labrd}}$($A$)\\
        $P$ and $Q$ initially empty;\\
      \For{$i =1:b $}{
        (a) $A_{i:m,i} =A_{i:m,i} - {P_{2(i-1)}Q_{2(i-1)}^\mathrm{T}}$; \CommentSty{//update $i$-th column (\textbf{gemv}$\times$\textbf{1})}\\
        (b) \CommentSty{//compute Householder reflector $P_i$ to eliminate below diagonal:}\\
        \quad $(\tau_i, v_i)=\texttt{larfg}(m-i, A_{i,i}, A_{i+1:m,i})$;\\
         \quad $y_i =\tau_i \left( A-P_{2(i-1)}Q_{2(i-1)}\right)^\mathrm{T} v_i$; \CommentSty{//(\texttt{gemv}, \textbf{gemv}$\times$\textbf{2})}\\
        
         \quad ${P}_{2i-1} = [{P}_{2(i-1)}, {v}_i]$, ${Q}_{2i-1} = [{Q}_{2(i-1)}, y_i]$; \CommentSty{//save $v_i$ and $y_i$}\\ 
         
        (c) $A_{i,i+1:n}=A_{i,i+1:n}-{P_{2i-1}Q_{2i-1,i+1:n}^\mathrm{T}}$; \CommentSty{//update $i$-th row  (\textbf{gemv}$\times$\textbf{1})}\\

        (d) \CommentSty{compute Householder reflector $Q_i$ to eliminate right of off-diagonal:}\\
        \quad $(\pi_i, u_i)=\texttt{larfg}(n-i-1, A_{i,i+1}, A_{i,i+2:n})$;\\ 
         \quad $x_i =  \pi_i \left( A-P_{2i-1}Q_{2i-1}\right)^\mathrm{T} u_i$; \CommentSty{//(\texttt{gemv}, \textbf{gemv}$\times$\textbf{2})}\\ 
         
         \quad ${Q}_{2i} = [{Q}_{2i-1}, {u}_i]$, ${P}_{2i} = [{P}_{2(i-1)}, {x}_i]$; \CommentSty{//save $u_i$ and $x_i$}\\ 
        }
        \textbf{return}$\left( P_{b+1:m,1:2b}, Q_{b+1:n,1:2b} \right)$;\\
   \textbf{end function}
   
\end{algorithm}

Algorithm \ref{alg:bi-diag-block} describes the pseudocode of our proposed blocked bidiagonalization procedure. 

\subsubsection{\textbf{Accelerating Bidiagonalization on GPU}}

\begin{wrapfigure}{r}{0.4\textwidth} 
    \centering
   \includegraphics[width=0.9\linewidth]{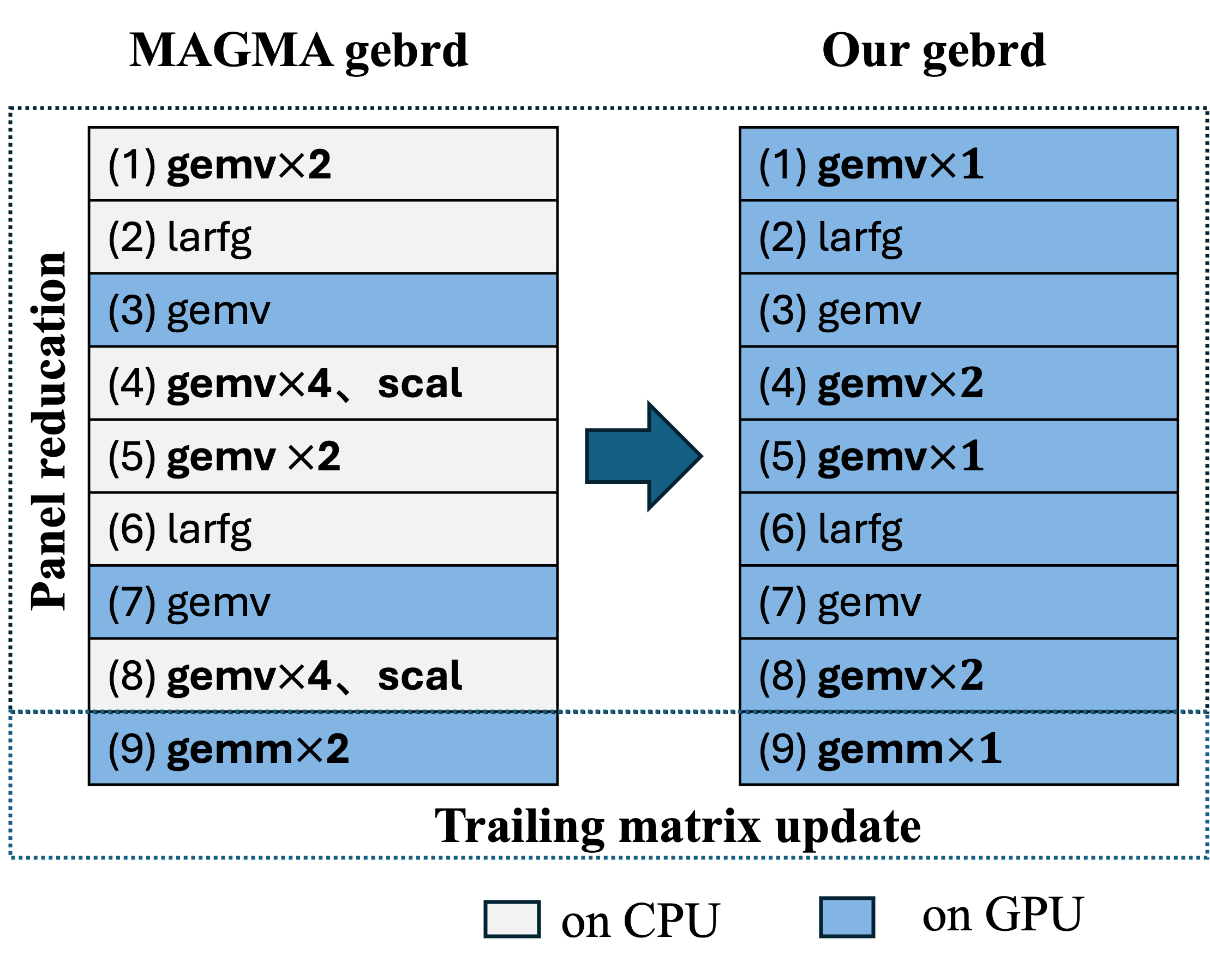}
    \caption{Comparison of MAGMA’s \texttt{gebrd} with our approach. Bolded operations highlight the differences from MAGMA. 
    }
    \label{fig:gebrd-magma-fast-operations}
\end{wrapfigure}

The primary computational cost of bidiagonalization lies in the trailing matrix-vector products (\texttt{gemv}) and trailing matrix 
updates (\texttt{gemm}).
Accordingly, MAGMA schedules these two 
operations on GPU, while the remaining computations are executed on CPU, as shown in Fig.~\ref{fig:gebrd-magma-fast-operations}. 
However,
this strategy incurs substantial CPU-GPU data transfers. 
Although data transfers and trailing-matrix multiplications are partially overlapped by panel-level computations in the algorithmic pipeline, their impact remains limited. 
This is mainly due to inherent inefficiencies in CPU-GPU communication and the fact that trailing matrix updates are not the dominant 
bottleneck.
In our method, both panel-level computations and trailing matrix updates are performed entirely  on GPU,
with their operations merged as described earlier. 
The changes in computational requirements compared to the MAGMA algorithm are highlighted in bold in Fig.~\ref{fig:gebrd-magma-fast-operations}. Specifically, the innovations in panel-level computations are as follows:
\begin{itemize}
    \item Step (1) reduces two \texttt{gemv} operations to one for updating the current column.
    \item Step (4) merges four \texttt{gemv} operations into two for computing $\bm{y}_i$, while integrating the \texttt{scal} operation into \texttt{gemv}.
    \item Step (5) similarly reduces two \texttt{gemv} operations to one for updating the current row.
    \item Step (8) performs the same merging as in Step (4), reducing four \texttt{gemv} operations to two for computing $\bm{x}_i$, while integrating the \texttt{scal} operation into \texttt{gemv}.
\end{itemize}
For the trailing matrix updates, step (9) combines two \texttt{gemm} operations into a single combined \texttt{gemm} operation.

\begin{wrapfigure}{r}{0.5\textwidth} 
        \centering
        \includegraphics[width=0.9\linewidth]{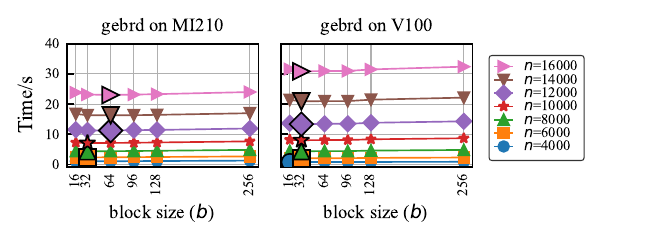}
        \caption{Tuning of our \texttt{gebrd} with varying block size ($b$).}
        \label{fig:mi210-v100-dgebrd-nb-time}
\end{wrapfigure}

\begin{figure}[ht]
    \centering
        
    \begin{subfigure}[t]
         {0.4\textwidth}
         \centering
          \centering\includegraphics[width=1\linewidth]{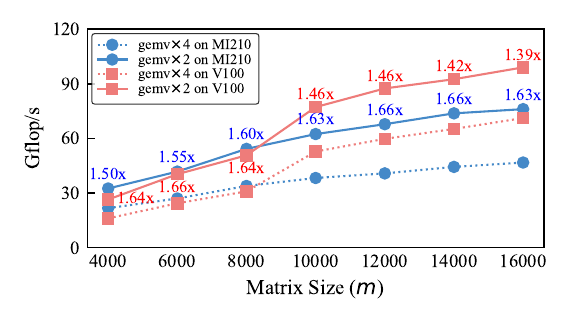}
        \caption{Merged \texttt{gemv}$\times$2 vs. non-merged \texttt{gemv}$\times$4.}
        \label{fig:mi210-v100-dgemv-flops}
        \end{subfigure}
        \begin{subfigure}[t]
         {0.4\textwidth}
         \centering
         \includegraphics[width=1\linewidth]{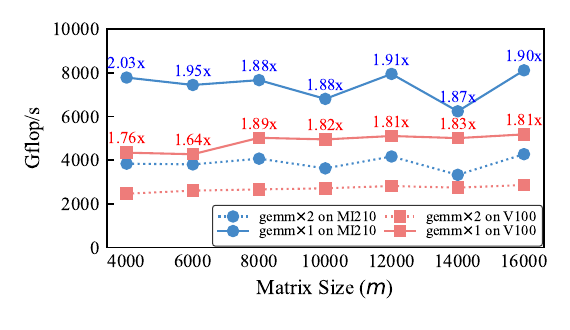}
        
        \caption {Merged \texttt{gemm}$\times$1 vs. non-merged \texttt{gemm}$\times$2.
        }\label{fig:mi210-v100-dgemm-flops}
    \end{subfigure}
    \caption{Performance comparison of merged vs. non-merged operations with speedup annotations for MI210 (blue) and V100 (red).}\label{fig:mi210-v100-dgemv-dgemm-flops}
\end{figure}

As depicted in Fig.~\ref{fig:mi210-v100-dgebrd-nb-time}, the block size ($b$) affects \texttt{gebrd} performance, with the optimal size indicated by a larger marker and employed throughout subsequent experiments.
In the panel-level factorization, we reduce the number of \texttt{gemv} operations for computing each of $\bm{y}$ and $\bm{x}$ from four to two.
Fig.~\ref{fig:mi210-v100-dgemv-flops} compares the performance of the original formulation
$\bm{\hat{x}}=(\bm{V}\bm{Y}^\mathrm{T}+\bm{X}\bm{U}^\mathrm{T})\bm{u}$, where $\bm{V},\bm{Y},\bm{X},\bm{U}\in  \mathbb{R}^{m\times32}$ (\texttt{gemv}$\times$4) 
against the merged  version $\bm{\hat{x}}= \bm{P}\bm{Q}^\mathrm{T}\bm{u}$, where $\bm{P} = [\bm{V} \, \bm{X}],\bm{Q} = [\bm{Y} \, \bm{U}]\in \mathbb{R}^{m \times 64}$
(\texttt{gemv}$\times$2).
Fig.~\ref{fig:mi210-v100-dgemm-flops} evaluates trailing matrix 
updates 
$\bm{A}=\bm{A}-\bm{V}\bm{Y}^\mathrm{T}-\bm{X}\bm{U}^\mathrm{T}$ (\texttt{gemm}$\times$2) 
versus the merged update $\bm{A}=\bm{A}-\bm{P}\bm{Q}^\mathrm{T}$ (\texttt{gemm}$\times$1). 
Speedup annotations for MI210 and V100 are indicated by blue and red numbers, respectively.
Fig.~\ref{fig:mi210-v100-dgemv-dgemm-flops} shows that the merged \texttt{gemv}$\times$2 and \texttt{gemm}$\times$1 achieve significant performance gains across all scales and platforms. 
Additionally, for $m > 8000$, \texttt{gemv} performance is higher on V100 than on MI210, while MI210 consistently outperforms V100 in \texttt{gemm} across all scales.

\begin{figure}[ht]
    \centering
           \includegraphics[width=0.9\linewidth]{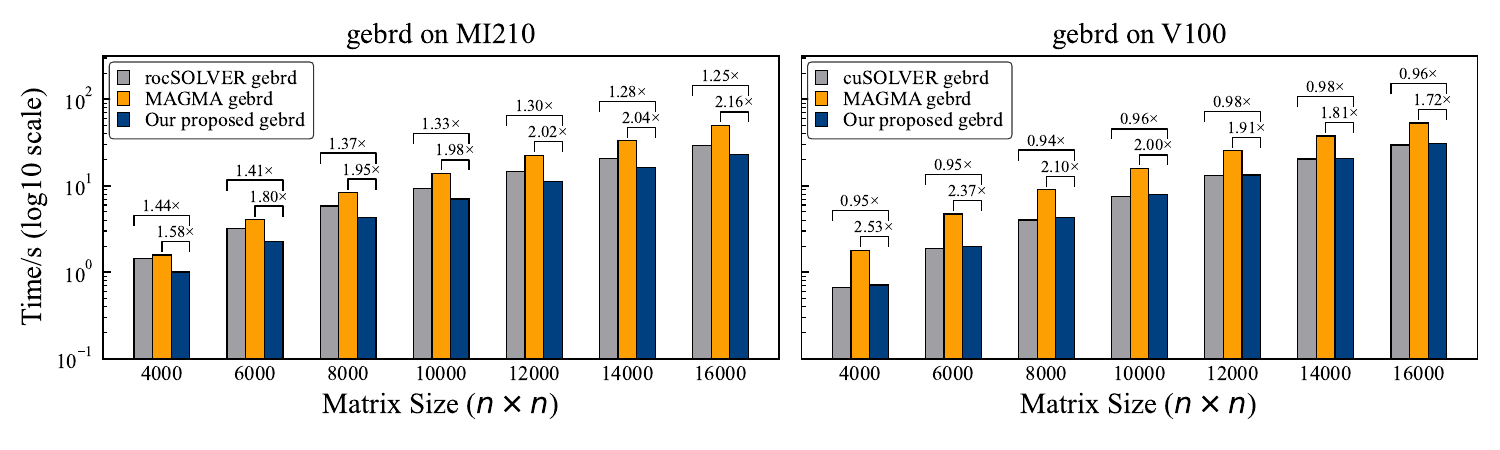}
        
        \caption{
        Performance comparison of \texttt{gebrd} implementations: rocSOLVER/cuSOLVER, MAGMA, and our proposed method. Speedups over rocSOLVER/cuSOLVER and MAGMA are annotated above the bars.
        }\label{fig:mi210-v100-dgebrd-roc-cu-ma-our-time}
\end{figure}

Fig.~\ref{fig:mi210-v100-dgebrd-roc-cu-ma-our-time} compares the performance of our proposed \texttt{gebrd} method against rocSOLVER/cuSOLVER, and MAGMA for square matrices on MI210 and V100 GPUs. 
The numerical values above bars represent the speedup achieved by our method over rocSOLVER/cuSOLVER and MAGMA.
As shown, our method consistently outperforms rocSOLVER and MAGMA across all tested matrices, achieving speedups of up to 1.44x over rocSOLVER and up to 2.16x and 2.53x over MAGMA on the MI210 and V100, respectively.
On V100, our method achieves performance comparable to cuSOLVER’s \texttt{gebrd}, reaching up to 98\% of its performance.

\subsection{Diagonalization}

\subsubsection{\textbf{BDC Algorithm}}
After the matrix is reduced to bidiagonal form, the BDC algorithm is employed to compute the SVD of the bidiagonal matrix $\bm{B}$, such that $\bm{B} = \bm{U} \bm{\Sigma} \bm{V}^\mathrm{T}$.
A brief introduction to the BDC algorithm is provided here; for more details, please refer to \citep{gu1995divide,jessup1994parallel,demmel1997applied}.
The BDC algorithm consists of three stages: (1) divide a big problem into smaller subproblems
recursively, (2) solve the small subproblems, and (3) conquer the solutions of the subproblems.
Let $\bm{B} \in \mathbb{R}^{n \times (n + 1)}$ be an upper bidiagonal matrix.
BDC first divides $\bm{B}$ into smaller submatrices:
$\bm{B}= \left[\begin{array}{cc}
       \bm{B}_1  & \\
       {\alpha_{k}} {\bm{e}_{k}}^\mathrm{T} & \beta_k \bm{e}_1^\mathrm{T}\\
       & \bm{B}_2
    \end{array}
    \right]$, 
where $\bm{B}_1 \in \mathbb{R}^{(k-1) \times k}$ and $\bm{B}_2 \in \mathbb{R}^{(n-k) \times (n-k+1)}$ are upper bidiagonal matrices. 
Typically, $k = \lfloor n/2 \rfloor$, and $\bm{e}_k$ denotes the $k$-th standard basis vector.
Assume the SVDs of $\bm{B}_1$ and $\bm{B}_2$ are given by
\begin{equation}
\label{eq:b1-b2-svd}
\bm{B}_i = 
\bm{W}_i \left[ \begin{matrix}
   \bm{D}_i & \bm{0}
\end{matrix}\right] \left[ \begin{matrix} \bm{Q}_i & \bm{q}_i \end{matrix} \right]^\mathrm{T},\ i=1,2,
\end{equation}
where $\bm{W}_i$ and $\left[ \bm{Q}_i \ \bm{q}_i \right]$ are orthonormal matrices, and $\bm{D}_i$ is a non-negative diagonal matrix.
For the base case, when the size of $\bm{B}_i$ is small enough, its SVD can be computed by QR iteration (called \texttt{lasdq} in LAPACK). 
To compute the SVD of matrix $\bm{B}$ from the SVDs of its submatrices $\bm{B}_1$ and $\bm{B}_2$, a technique known as deflation is employed. Deflation identifies and isolates the converged singular values along with their corresponding singular vectors, thereby reducing the remaining problem size. After deflation, the matrix is restructured as follows:
\begin{equation}\label{eq:bdc-B}
    \bm{B} = \left[ \tilde{\bm{W}} \quad \bm{W}_d\right]
    \left[\begin{matrix}
        \bm{M} & 0 & 0\\
        0 & \bm{\Omega}_d & 0\\
    \end{matrix} \right]
    \left[ \tilde{\bm{Q}} \quad \bm{Q}_d \quad \bm{q}\right]^\mathrm{T},
\end{equation}
where $\bm{\Omega}_d$ represents the deflated singular values, and $\bm{W}_d$ and $\bm{Q}_d$ are the deflated singular vectors, $\bm{M}$ is a matrix with a special structure, which will be introduced in the following content. See \citep{gu1995divide} for further details. Additionally,
\begin{equation}\label{eq:bdc-W-Q}
    \tilde{\bm{W}}=\left[ \begin{matrix}
    0 & \tilde{\bm{W}}_{0,1} & \tilde{\bm{W}}_{1} & 0\\
    1 & 0 & 0 & 0\\
    0 & \tilde{\bm{W}}_{0,2} & 0 & \tilde{\bm{W}}_2
    \end{matrix} \right] \ \text{and} \ \tilde{\bm{Q}}=\left[
    \begin{matrix}
     \tilde{\bm{Q}}_{0,1} & \tilde{\bm{Q}}_{1} & 0\\
    \tilde{\bm{Q}}_{0,2} & 0 & \tilde{\bm{Q}}_{2}\\
    \end{matrix}
     \right], 
\end{equation}
where $\tilde{\bm{W}}_i$, $\tilde{\bm{W}}_{0,i}$, $\tilde{\bm{Q}}_i$, and $\tilde{\bm{Q}}_{0,i}$ are derived from ${\bm{W}}_i$, ${\bm{Q}}_i$  and ${\bm{q}}_i$ through the deflation process. 
Let $\bm{U}\bm{\Omega}\bm{V}^\mathrm{T}$ be the SVD of $\bm{M}$.
Substituting this into \eqref{eq:bdc-B} yields $\bm{B}$:
\begin{equation}
\label{eq:bdc-B-svd}
    \bm{B} = \left[ \tilde{\bm{W}}\bm{U} \quad \bm{W}_d\right]
    \left[\begin{matrix}
        \bm{\Omega} & 0 & 0\\
        0 & \bm{\Omega}_d & 0\\
    \end{matrix} \right]
    \left[ \tilde{\bm{Q}}\bm{V} \quad \bm{Q}_d \quad q\right]^\mathrm{T}.
\end{equation}
By exploiting the block structure in \eqref{eq:bdc-W-Q}, the updated singular vectors $\tilde{\bm{W}}$ and $\tilde{\bm{Q}}$ can each be computed using three matrix-matrix multiplications (\texttt{gemm} $\times$ 3).
\begin{equation}
\label{eq:bdc-Mv2Bv-u-vt}
   \tilde{\bm{W}}\bm{U}=\left[ \begin{matrix}
        \tilde{\bm{W}}_{0,1} \bm{U}_{0} + \tilde{\bm{W}}_{1}\bm{U}_{1} \\
        \bm{u}_{0}^\mathrm{T}\\
        \tilde{\bm{W}}_{0,2} \bm{U}_{0} + \tilde{\bm{W}}_{2}\bm{U}_{2}
     \end{matrix} \right],
     \tilde{\bm{Q}}\bm{V}=\left[ \begin{matrix}
      \tilde{\bm{Q}}_{0,1} \bm{V}_{0} + \tilde{\bm{Q}}_{1}\bm{V}_{1} \\
      \tilde{\bm{Q}}_{0,2} \bm{V}_{0} + \tilde{\bm{Q}}_{2}\bm{V}_{2}
     \end{matrix} \right],  \ \text{where} \ 
     \bm{U}=\left[ \begin{matrix}
         \bm{u}_{0}^\mathrm{T}\\
         \bm{U}_{0}\\
         \bm{U}_{1}\\
         \bm{U}_{2}
     \end{matrix} \right] \ \text{and} \ 
     \bm{V}=\left[ \begin{matrix}
         \bm{V}_{0}\\
         \bm{V}_{1}\\
         \bm{V}_{2}
     \end{matrix} \right].
\end{equation}


Next, we introduce the SVD of the matrix $\bm{M}$. The matrix $\bm{M}$ possesses a special structure

\begin{equation}\label{eq:diag-one-side-M}
   \bm{M}=\left [ \begin{matrix}
    z_1 &z_2 &\dots &z_N\\
     &  d_2 &  & \\
     &  & \ddots & \\
     &  &  & d_N
\end{matrix} \right ],
\end{equation}
where $N$ is the number of non-deflated singular values.
Let $\bm{D}=\texttt{diag}(d_1, d_2, \cdots, d_N)$, with $d_1\equiv0$; and $\bm{z}=(z_1, z_2, \cdots, z_N)^\mathrm{T}$.
The singular values ${\{\omega_i\}}_{i=1}^{N}$ of  $\bm{M}$ are the roots of the secular equation, 
\begin{equation}
\label{eq:M-svd-val}
    f(\omega_i )=1+{\textstyle \sum_{j=1}^{N}} \frac{z_j^2}{d_j^2-\omega_i^2}=0.
\end{equation}
Although the computed singular values have  highly relative accuracy, 
small approximation errors may cause a loss of orthogonality in the computed singular vectors. 
To address this,
Gu and Eisenstat \citep{gu1995divide} propose computing a new matrix $\tilde{\bm{M}}$, structured similarly to $\bm{M}$, for which the computed ${\{\tilde{\omega}_i\}}_{i=1}^{N}$ are the exact singular values, with 
\begin{equation}\label{eq:bdc-zi}
   |\bm{\tilde{z}}_i| = \sqrt{\left( \tilde{\omega}_N^2 -d_i^2 \right) \prod_{k=1}^{i-1} \frac{\tilde{\omega}_k^2-d_i^2}{d_k^2-d_i^2} \prod_{k=i}^{N-1} \frac{\tilde{\omega}_k^2-d_i^2}{d_{k+1}^2-d_i^2}},  
\end{equation}
where the sign of $\tilde{z}_i$ can be chosen arbitrarily.

The left and right singular vectors of $\tilde{\bm{M}}$ are then computed as follows:
  \begin{equation}\label{eq:M-svd-vec}
   \begin{split}
        & \bm{v}_i=
        \left [
        \frac{\tilde{z}_1}{d_1^2-\tilde{\omega}_i^2},
        \frac{\tilde{z}_2}{d_2^2-\tilde{\omega}_i^2},
        \dots,
        \frac{\tilde{z}_N}{d_N^2-\tilde{\omega}_i^2}  \right ]^\mathrm{T}
        =[v_{i1}, v_{i2}, \cdots, v_{iN}]^\mathrm{T}, \
        \bm{v}_i =\frac{\bm{v}_i}{\left \| \bm{v}_i \right \|_2}, \
        \\
        & 
        \bm{u}_i 
        = \left [-1, \frac{d_2 \tilde{z}_2}{d_2^2-\tilde{\omega}_i^2},\dots,\frac{d_N\tilde{z}_N}{d_N^2-\tilde{\omega}_i^2}  \right ]^\mathrm{T} 
        = \left [-1, d_2 v_{i2},  \cdots, d_N v_{iN}   \right ]^\mathrm{T}, 
        \ \bm{u}_i=\frac{\bm{u}_i}{\left \|\bm{u}_i\right \|_2}.
    \end{split}
\end{equation}

Further, the matrix $\bm{M}$ needs to be satisfied 
\begin{equation}\label{eq:M-condition}
    |d_i-d_j| \geq \varepsilon \left \| M \right \|_2 \ \text{for} \ i \neq j,
    \quad |z_i| \geq \varepsilon \left \| M \right \|_2,
\end{equation}
where $\varepsilon$ is a small multiple of the machine precision. If it is not satisfied, the matrix $\bm{M}$ has to be deflated before computing its SVD (called \texttt{lasd2}). 
Here, we briefly introduce the deflation proces; for details, see \citep{gu1995divide}. We illustrate the reduction for $N=3$.
There are two scenarios in which deflation can occur:
\begin{enumerate}[left=0pt]
    \item Small $\bm{z}$-component deflation.
    \begin{itemize}[left=0pt]
        \item{ If $|z_1|<\varepsilon \left \| M \right \|_2$, then set $|z_1|=\varepsilon \left \| M \right \|_2$: 
        $\bm{M}=\left[ \begin{matrix}
            z_1 & z_2 & z_3\\
             & d_2& \\
             & & d_3\\
        \end{matrix}\right] = \left[ \begin{matrix}
            \varepsilon \left \| M \right \|_2 & z_2 & z_3\\
             & d_2& \\
             & & d_3\\
        \end{matrix}\right] + O(\varepsilon \left \| M \right \|_2)$.   
        }
        \item {If $|z_i| < \varepsilon \left \| M \right \|_2  for i \geq 2$, then set $z_i = 0$ (e.g., for $i = 3$):
            $\bm{M}=\left[ \begin{matrix}
                z_1 & z_2 & z_3\\
                 & d_2& \\
                 & & d_3\\
            \end{matrix}\right] =  \left[ \begin{matrix}
                z_1 & z_2 & 0\\
                 & d_2& \\
                 & & d_3\\
            \end{matrix}\right] + O(\varepsilon \left \| M \right \|_2)$.  
        }
    \end{itemize}
   
    \item Close singular value deflation.
    \begin{itemize}
        \item {Suppose $|d_1 - d_i| < \varepsilon \left \| M \right \|_2$. Let $r=\sqrt{z_1^2 + z_i^2}, c = z_1/r$ and $s = z_i/r$. Then set $d_i = 0$, and apply a Givens rotation to e zero out $z_i$ (e.g., for $i = 3$):
            \begin{displaymath}
              \bm{M} \bm{G}^\mathrm{T} =
                \left[ \begin{matrix}
                    z_1 & z_2 & z_3\\
                     & d_2& \\
                     & & 0\\
                \end{matrix}\right]
                 \left[ \begin{matrix}
                    c &  & -s\\
                     & 1 & \\
                    s & & c\\
                \end{matrix}\right]
                + O(\varepsilon \left \| M \right \|_2) =
                \left[ \begin{matrix}
                    r & z_2 & 0\\
                     & d_2 & \\
                     & & 0\\
                \end{matrix}\right] + O(\varepsilon \left \| M \right \|_2).  
            \end{displaymath}
            }
            
        \item {Suppose $|d_i - d_j| < \varepsilon \left \| M \right \|_2, 
 i, j \geq 2$. Let $r=\sqrt{z_i^2 + z_j^2}, c = z_j/r$ and $s = z_i/r$. Then replace $d_j$ with $d_i$, and perform a Givens rotation to zero out $z_i$ (e.g., for $i = 3, j = 2$):
        \begin{displaymath}  
              \bm{G}\bm{M}\bm{G}^\mathrm{T} = 
                \left[ \begin{matrix}
                    1 &   &  \\
                     & c & s \\
                     & -s & c\\
                \end{matrix}\right]
                \left[ \begin{matrix}
                    z_1 & z_2 & z_3\\
                     & d_3 & \\
                     & & d_3\\
                \end{matrix}\right]
                \left[ \begin{matrix}
                    1 &   &  \\
                     & c & -s \\
                     & s & c\\
                \end{matrix}\right] + O(\varepsilon \left \| M \right \|_2)   =
                 \left[ \begin{matrix}
                    z_1 & r  & 0\\
                     &  d_3 &  \\
                     &  & d_3 \\
                \end{matrix}\right] + O(\varepsilon \left \| M \right \|_2).  
            \end{displaymath}
        }
    \end{itemize}
\end{enumerate}    
Clearly, by applying the above techniques and rearranging the diagonal elements, we can obtain two orthogonal matrices, $\bm{P}$ and $\bm{Q}$, such that
\begin{displaymath}
    \bm{P}\bm{M}\bm{Q}=\left[\begin{matrix}
        \bm{M}_1 & \\
        & \bm{D}
    \end{matrix}\right] + O(\varepsilon \left \| M \right \|_2),
\end{displaymath}
where $\bm{M}_1$, which has the same structure as $\bm{M}$ but with a smaller dimension, satisfies condition $\eqref{eq:M-condition}$. $\bm{D}$ is a diagonal matrix with non-negative entries. 
Therefore, we only need to apply the previously introduced methods to $\bm{M}_1$.

Therefore, the BDC algorithm can be outlined in Algorithm \ref{alg:bdc}.
\begin{algorithm}[ht]\small
\caption{A Pseudocode of the BDC Algorithm}\label{alg:bdc}
    (1) For the nodes on bottom level of the tree, solve subproblems by QR iteration; \CommentSty{//(\texttt{lasdq})}\\
    (2) Conquer each subproblem bottom-up;\CommentSty{//(\texttt{lasd1})}\\
    \quad $\textbf{for} \ i=H \ to \ 1$ \ \CommentSty{//$H$ is the hight of the tree}\\
        \ \quad a) Deflate singular values; \CommentSty{//(\texttt{lasd2})}\\
        \ \quad b) Solve Secular equation and update singular vectors; \CommentSty{//(\texttt{lasd3})}\\
    \quad $\textbf{endfor}$\\
    (3)  Sort the singular values and corresponding singular vectors; 
\end{algorithm}

\subsubsection{\textbf{Accelerating BDC on GPU}}
\begin{figure}[ht]
    \centering
     \begin{minipage}{0.41\textwidth}
     \includegraphics[width=1\linewidth]{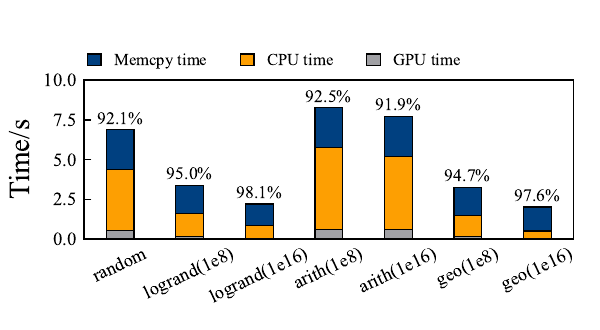}
    \caption{Profiling of \cite{gates2018accelerating}' \texttt{lasd3} at the root level for $n$=20000 on MI210, with  percentage of CPU+Memcpy time is annotated above each set of bars.}
    \label{fig:mi210-v1-lasd3-root-part-2w}
    \end{minipage}
    \quad
     \begin{minipage}{0.56\textwidth}
     \includegraphics[width=1\linewidth]{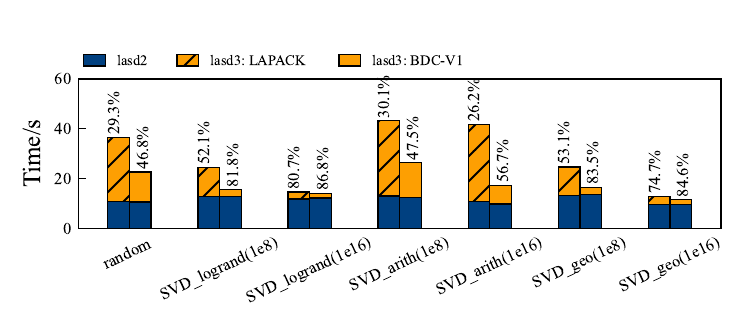}
    \caption{Profiling of
    BDC at the root level for $n$=20000 on MI210 with percentage of \texttt{lasd2} annotated above each set of bars.
    }
    \label{fig:mi210-dbdsdc-la-v1-root-part}
     \end{minipage}
\end{figure}

There are several potential sources of parallelism in the BDC algorithm. For example, in the recursion tree, each subproblem is independent. Most of the time is spent in the merge step, in particular, in the matrix multiplies $\tilde{\bm{W}}\bm{U}$ and $\tilde{\bm{Q}}\bm{V}$ by \eqref{eq:bdc-Mv2Bv-u-vt},
respectively. 
Further, most of the cost is at the higher levels of the recursion tree, near the root node, as the matrices get larger, rather than in the leaf nodes. 
In the method proposed by \cite{gates2018accelerating} (called BDC-V1), the focus is on the merge step (\texttt{lasd3}), with only the \texttt{gemm} operations—associated with singular vector updates in \eqref{eq:bdc-Mv2Bv-u-vt}—to the GPU. The remaining CPU-based computations and CPU-GPU data transfers, despite partial GPU overlap, become the primary bottleneck as GPU-accelerated \texttt{gemm} efficiency increases, as shown in Fig.~\ref{fig:mi210-v1-lasd3-root-part-2w}.
Furthermore, the time spent on \texttt{lasd2} in the BDC algorithm is also substantial. As shown in Fig.~\ref{fig:mi210-dbdsdc-la-v1-root-part}, a comparison of LAPACK and BDC-V1 across four matrix types with varying condition numbers highlights \texttt{lasd2}’s substantial contribution, underscoring the need for targeted optimization.
Consequently, we present a comprehensive description of our GPU-based approach for optimizing the two primary components of the BDC algorithm: \texttt{lasd2} and \texttt{lasd3}.

\begin{wrapfigure}{r}{0.4\textwidth}
    \centering
    \includegraphics[width=1\linewidth]{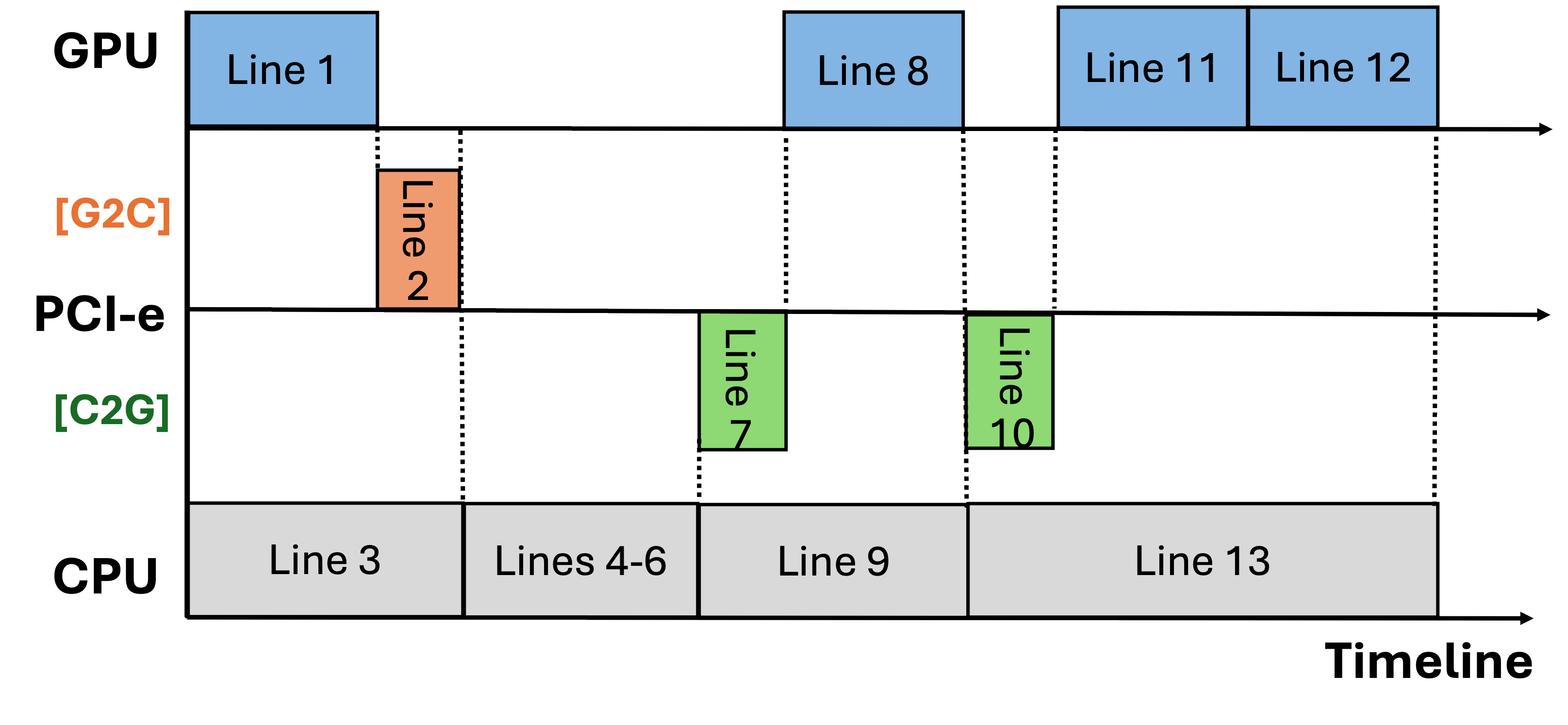}
    \caption{Timeline of Algorithm \ref{alg:bdc-lasd2}: GPU-based \texttt{lasd2}. 
    }
    \label{fig:lasd2-gpu-timeline}
\end{wrapfigure}

(1) The GPU-based \texttt{lasd2} deflation subroutine is presented in Algorithm \ref{alg:bdc-lasd2}, with its execution timeline shown in Fig.~\ref{fig:lasd2-gpu-timeline}.
First, the $\bm{z}$-vector is computed on GPU by multiplying the singular
vector matrix $\bm{V}$ with the diagonal and 
off-diagonal elements of the added row. 
Once computed, it is transferred to CPU for subsequent processing (line 2).
Singular value sorting (line 3) on CPU can overlap with the $\bm{z}$-vector computation and data transfer (lines 1$\sim$2).
Over $n{-}1$ iterations, CPU handles $\bm{z}$-component deflation and Givens rotations for close singular values, while GPU updates of $\bm{U}$ and $\bm{V}$ overlap with subsequent CPU operations in the next iteration (lines 5$\sim$6).
After deflation, singular values are sorted on CPU (line 8), overlapping with the application of rotations on GPU from the last iteration (line 7). 
The corresponding index information is then transferred to the GPU, where the singular vectors are permuted based on these indices.
Finally, the deflated singular vectors are copied back to their respective positions in the matrices $\bm{U}$ and $\bm{V}$ on GPU (line 11), while the deflated singular values are synchronized and copied to $\bm{D}$ on CPU (line 12).
Additionally, the sorting process in the step (3) of Algorithm \ref{alg:bdc} is similar to that in lines 8$\sim$10 of Algorithm \ref{alg:bdc-lasd2}.

\begin{algorithm}[ht]\small
    \caption{A Pseudocode of Our GPU-based \texttt{lasd2} Subroutine}\label{alg:bdc-lasd2}
    
   {[on GPU]} Generate the $\bm{z}$-vector;\\
    {[G2C]} Transfer $\bm{z}$ from GPU $\longrightarrow$ CPU;\\
    {[on CPU]} Sort singular values into increasing order;
    //\CommentSty{can overlap with lines 1$\sim$2}\\
     \For{$i=2:n $}{ 
        {[on CPU]} Deflate due to small $\bm{z}$-vector component;\\
        {[on CPU]} For close singular values, compute the Givens rotation;\\
        {[on GPU]} Apply the Givens rotations to $\bm{U}$ and $\bm{V}$; //\CommentSty{can overlap with lines 5$\sim$6 of the next iteration}\\
    }
    {[on CPU]} Sort singular values and restore indices; //\CommentSty{can overlap with line 7 of the last iteration}\\
    {[C2G]} Transfer indices from CPU $\longrightarrow$ GPU; //\CommentSty{can overlap with line 7 of the last iteration}\\
    {[on GPU]} Permute singular vectors according to indices;\\
    {[on GPU]} Copy deflated singular vectors to the back of $\bm{U}$ and $\bm{V}$;\\
    {[on CPU]} Copy deflated singular values to the back of $\bm{D}$; //\CommentSty{can overlap with lines 9$\sim$11}
    \\
\end{algorithm}

\begin{figure}[ht]
   \centering
   \includegraphics[width=0.7\linewidth]{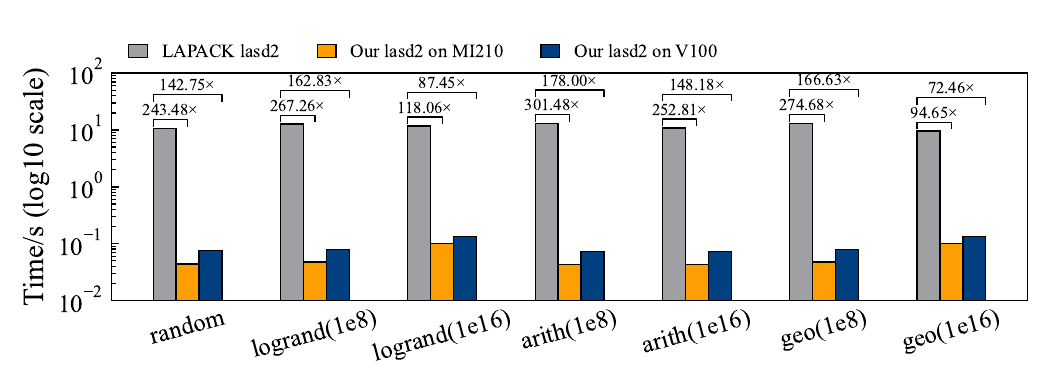}
    \caption{
    Performance comparison of \texttt{lasd2} at the root node: LAPACK vs. our GPU-based method for $n = 20000$. Speedups on MI210 and V100 are annotated above the bars.
    }
    \label{fig:mi210-v100-dlasd2-la-v2-2w}
\end{figure}

Fig.~\ref{fig:mi210-v100-dlasd2-la-v2-2w} compares the performance of LAPACK and our GPU-based \texttt{lasd2} method at the root node across various matrix types with $n$=20000. The achieved speedups of our \texttt{lasd2} implementation on both MI210 and V100, relative to LAPACK, are annotated above bars. 
The results clearly show that our \texttt{lasd2} method delivers substantial performance improvements across all matrix types, with particularly notable gains on MI210. 
Our GPU-based \texttt{lasd2} method efficiently manages deflation during the SVD update by utilizing both CPU and GPU, avoiding matrix-level data transfers, and overlapping CPU tasks, GPU kernels, and CPU-GPU communication.

(2) For our GPU-based \texttt{lasd3} merge subroutine method, as detailed in Algorithm \ref{alg:bdc-lasd3}, line 5 can be efficiently executed using GPU-based BLAS library functions.
\begin{algorithm}[ht]\small
    \caption{A Pseudocode of Our GPU-based \texttt{lasd3} Subroutine }\label{alg:bdc-lasd3}
    \textbf{parallel}
    \For{$i = 1$ \textbf{to} $N$}{ 
        compute $\tilde{\omega}_i $ by solving secular \eqref{eq:M-svd-val}; //(\texttt{lasd4})
    }
    \textbf{[C2G]} copy ${\{d_i\}}_{i=1}^{N}$ and ${\{\tilde{\omega}\}}_{i=1}^{N}$ from CPU $\rightarrow$ GPU;\\
    \textbf{[on GPU]} compute $\bm{V}$ and $\bm{U}$ by \eqref{eq:bdc-zi} and \eqref{eq:M-svd-vec};\\
      \textbf{[on GPU]} compute $\bm{U} = \bm{Q} \bm{U}$ and $\bm{V} = \bm{W} \bm{V}$ using \eqref{eq:bdc-Mv2Bv-u-vt}, with \texttt{gemm} $\times$ 3 for each;
\end{algorithm}
In line 4 of Algorithm~\ref{alg:bdc-lasd3}, the computations of $\bm{U}$ and $\bm{V}$ can be fused into a single GPU kernel for improved efficiency.
In the computation of $\tilde{z}_i$ in \eqref{eq:bdc-zi}, each thread-$j$ within block-$i$ computes its local contribution, denoted as $\tilde{z}_{ij}$, and stores it in a register.
Leveraging registers reduces memory bank conflicts and enhances hardware utilization.
Subsequently, a warp-shuffle multiplication reduction is performed within each block using warp-level shuffle instructions (\texttt{\_shfl\_down}) to compute  $\tilde{z}_i$. This allows for direct data exchange among threads within the same warp, significantly reducing latency and improving performance.
Once $\tilde{z}_i$ is computed, it is used in \eqref{eq:M-svd-vec} to update the corresponding columns of the singular vector matrices $\bm{U}$ and $\bm{V}$ associated with block-$i$.


\begin{figure}[ht]
    \centering
    \includegraphics[width=0.72\linewidth]{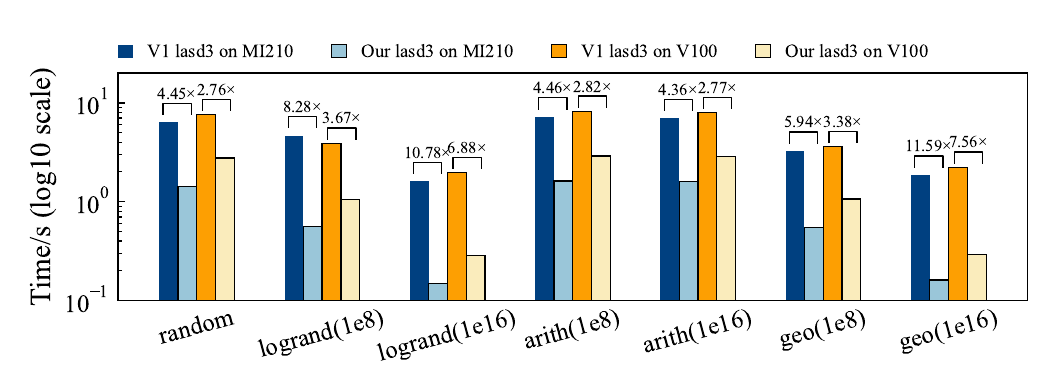}
    \caption{Performance comparison of \texttt{lasd3} at root node: BDC-V1 vs. our method for $n=20000$ with speedup annotations above bars for MI210 and V100.
     }
    \label{fig:mi210-v100-dlasd3-v1-v2-2w}
\end{figure}
\begin{figure}[ht]
    \centering
    \includegraphics[width=1\linewidth]{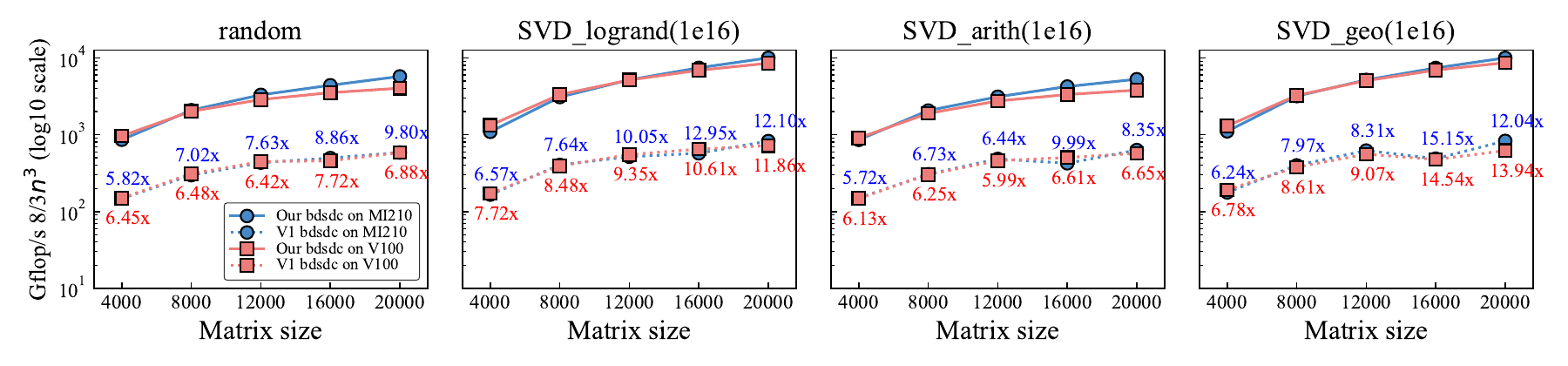}
    \caption{Performance comparison of \texttt{bdsdc}: BDC-V1 and our method on MI210 and V100.
    }\label{fig:dc-mkl-v1-v2-gflops-v100}
\end{figure}

Fig.~\ref{fig:mi210-v100-dlasd3-v1-v2-2w} shows a performance comparison of the \texttt{lasd3} routine at the root node between BDC-V1 and our proposed method for various matrix types ($n$=20000) on both MI210 and V100.
Speedups over BDC-V1 are shown above the bars.
As shown, our \texttt{lasd3} implementation achieves substantial performance improvements over BDC-V1 on both MI210 and V100,
with particularly pronounced improvements on MI210 due to its superior BLAS3 capabilities.
The size of leaf nodes impacts the performance of \texttt{bdsdc}. In our experiments, a leaf node size of 32 achieved the optimal performance.
Fig.~\ref{fig:dc-mkl-v1-v2-gflops-v100} shows the performance of the BDC algorithm across four matrix types. Assuming BDC requires approximately $\frac{8}{3}n^3$ operations, the actual operation count may be reduced due to deflation.
The speedups of our proposed \texttt{bdsdc} method over BDC-V1 on MI210 and V100 are indicated by the blue and red values along the dashed lines in Fig.~\ref{fig:dc-mkl-v1-v2-gflops-v100}. 
As shown, our \texttt{bdsdc} achieves substantial performance enhancements over BDC-V1 across all matrix types and sizes on both MI210 and V100 GPUs, reaching up to 12.04x and 13.94x, respectively.

\subsection{Initial QR Factorization of TS Matrix and Singular Vector Back-transformations}
\subsubsection{\textbf{Algorithm}}
If $m \gg n$, a more efficient strategy is to first compute a QR factorization $\bm{A} = \bm{Q} \bm{R}$, followed by an SVD of the smaller matrix $\bm{R} = \bm{U}_0 \bm{\Sigma} \bm{V}_0^\mathrm{T}$. 
The most widely used method for QR factorization is based on Householder transformations.
In the blocked Householder QR algorithm using the CWY transform, as implemented in LAPACK’s \texttt{geqrf} routine, each iteration consists of three steps:
\begin{enumerate}
    \item A panel of $b$ columns (with $b < n$) is factored using Householder transformations $\bm{H}_i, \dots, \bm{H}_{i+b}$, where each $\bm{H}_i = \bm{I} - \tau_i \bm{y}_i \bm{y}_i^{\mathrm{T}}$ (\texttt{geqr2}).
    \item A triangular matrix $\bm{T} \in \mathbb{R}^{b\times b}$ is 
    constructed from the inner products of the  reflectors in the panel (\texttt{larft}).
    \item The trailing matrix $\bm{A}_{\text{t}}$
    is updated by applying  $\bm{Y}$ and $\bm{T}$ from the left (\texttt{larfb}): 
\begin{equation}\label{eq:larfb}
     \bm{A}_{\text{t}}
    =  (\bm{I} - \bm{Y} \bm{T} \bm{Y}^\mathrm{T}) \bm{A}_{\text{t}}=\bm{A}_{\text{t}}-\bm{Y} \bm{T} \bm{Y}^\mathrm{T}\bm{A}_{\text{t}}, 
    \ \text{where} \ \bm{Y}=[\bm{y}_i, \cdots, \bm{y}_{i+b}].
\end{equation}
\end{enumerate}
After this update, the algorithm advances to the next panel and repeats until all columns are processed.
For TS SVD, the matrix $\bm{Q} \in \mathbb{R}^{m \times m}$ from the QR factorization should be generated. 
It can be computed via the CWY transform (called \texttt{orgqr}) as:
\begin{equation}
  \bm{Q} =  \left({\textstyle \prod_{i=1}^{n}}\bm{H}_i \right)  \bm{I} = {\textstyle \prod_{k=1}^{\lceil n/b \rceil}}(\bm{I} - \bm{Y}_k\bm{T}_k \bm{Y}_k^\mathrm{T})\bm{I}, 
    \label{eq:orgqr-q}
\end{equation}
where $\bm{H}_i$ are Householder reflectors, and $\bm{Y}_k$, $\bm{T}_k$ are the block representations from the $k$-th panel.
The generation of  $\bm{Q}$ clearly involves  two key steps: the construction of the triangular matrix $\bm{T}_k$  (\texttt{larft}) and  the update of the traling matrix (\texttt{larfb}).

For the back transformations of left and right singular vectors, $\bm{U} = \bm{U}_1 \bm{U}_2$ and $\bm{V}^\mathrm{T} = \bm{V}_2^\mathrm{T} \bm{V}_1^\mathrm{T} $. 
Here, $\bm{U}_2$ and $\bm{V}_2$  represent the singular vectors of the bidiagonal matrix.
The matrices $\bm{U}_1$ and $\bm{V}_1$, constructed as shown in \eqref{eq:gebrd-house}, denote the product of the column and row Householder reflectors obtained during the bidiagonalization process, respectively. Consequently,
\begin{equation}
    \bm{U} = \bm{U}_1 \bm{U}_2 = \left({\textstyle \prod_{i=1}^{n}}\bm{H}_i \right) \bm{U}_2
   \ \text{and} \ 
   \bm{V}^{\mathrm{T}} = \bm{V}^{\mathrm{T}}_2 \bm{V}^{\mathrm{T}}_1  =  \bm{V}^{\mathrm{T}}_2 \left({\textstyle \prod_{i=1}^{n}}\bm{G}_i \right)^{\mathrm{T}}.
\end{equation}
Clearly,
multiplying a matrix by $\bm{U}_1$ is  
similar to multiplying it by $\bm{Q}$ from a QR factorization, as 
in \eqref{eq:orgqr-q}  (called \texttt{ormqr}). 
Similarly, multiplying a matrix by $\bm{V}_1$ is equivalent to multiplying it by $\bm{Q}$ from an LQ factorization (called \texttt{ormlq}).


\subsubsection{\textbf{QR factorization on GPU}}
In the MAGMA method, the blocked Householder approach reduces overhead by delegating panel factorization to the CPU while overlapping it with trailing matrix updates on GPU. However, due to the GPU's superior computational capacity, the trailing matrix update is not the primary bottleneck; rather, the CPU-based panel computation and CPU-GPU data transfers become the primary limiting factors.
In our method, we have offloaded the panel-level computations to GPU and applied targeted optimizations to enhance performance. 

For the triangular factor $\bm{T}$, which can be constructed by
\begin{equation}
\label{eq:larft-1}
   \bm{T}_{i} =\left[ \begin{matrix}
        \bm{T}_{i-1} & \bm{z}_{i}\\
        0       & \tau_{i}
    \end{matrix} \right], \bm{z}_{i} = \bm{T}_{i} \left(-\tau_{i}\bm{Y}_{i-1}^\mathrm{T} \bm{y}_{i} \right), \bm{T}_0 = [], 1\leq i \leq b,
\end{equation}
where $\bm{Y}_{i-1}=[\bm{y}_1,\bm{y}_2,\cdots,\bm{y}_{i-1}]$.
Clearly, this process consists of $(b-1)$ iterations, each involving two BLAS2 operations:
\begin{align}
    \bm{w}_{i}=-\tau_{i} \bm{Y}_{i-1}^\mathrm{T} \bm{y}_{i} \qquad (\texttt{gemv})\\
    \bm{z}_{i} = \bm{T}_{i} \bm{w}_{i} \qquad (\texttt{trmv})
\end{align}
Unlike the standard CWY transform used in LAPACK and MAGMA, our GPU-based approach adopts the modified CWY transform similar to \citep{puglisi1992modification} to construct $\bm{T}^{-1}$:
\begin{equation}\label{eq:inv_t}
    \bm{T}^{-1}(i,j)=\left\{\begin{matrix}
{\bm{y}_i^{\rm{T}} \bm{y}_i}, & i>j\\
\frac{\bm{y}_i^{\rm{T}} \bm{y}_i}{2}, & i=j
\end{matrix}\right.
\end{equation}
Since $\tau_i=\frac{2}{\left \|\bm{y}_i \right \|_2^2}$, substituting it into \eqref{eq:inv_t} yields $\bm{T}^{-1}(i,i)=\frac{1}{\tau_i}$. Therefore, $\bm{T}^{-1}$ can be constructed as:
\begin{equation}\label{eq:larft_update}
    \bm{T}^{-1}=  \bm{Y}_{b}^\mathrm{T}  \bm{Y}_{b}
   \qquad (\texttt{syrk}/\texttt{gemm})
\end{equation}
\begin{equation}
    \texttt{diag}(\bm{T}^{-1})=(\frac{1}{\tau_1}, \frac{1}{\tau_2},\cdots,\frac{1}{\tau_b})
\end{equation}
While \texttt{syrk} is mathematically appropriate for symmetric updates, we use \texttt{gemm} in \eqref{eq:larft_update} for its superior performance and better optimization in vendor libraries such as rocBLAS and cuBLAS.

The trailing matrix update, as given in \eqref{eq:larfb}, 
is reformulated using $\bm{T}^{-1}$ as:

\begin{align}
  \bm{Z}=\bm{Y}^{\rm{T}} \bm{A}_t \qquad (\texttt{gemm})\\
  \bm{Z}=(\bm{T}^{-1})^{-1}\bm{Z} \qquad (\texttt{trsm})\\
  \bm{A}_t=\bm{A}_t-\bm{Y}\bm{Z} \qquad (\texttt{gemm})
\end{align}
This modified CWY formulation relies exclusively on compute-bound BLAS3 operations, substantially increasing arithmetic intensity and making it highly efficient for GPU execution.

\begin{figure}[ht]
    \centering
    \includegraphics[width=0.9\linewidth]{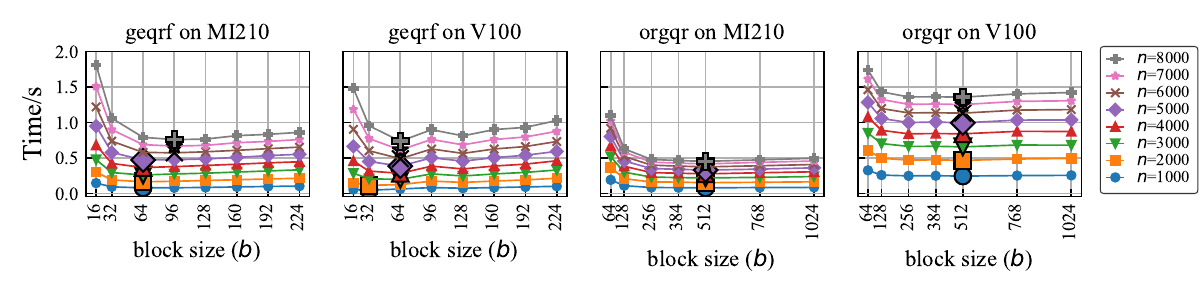}
    \caption{Tuning of our \texttt{geqrf} and \texttt{orgqr} with varying block size ($b$) for $m$=20000.}
    \label{fig:mi210-v100-dgeqrf-dorgqr-nb-time}
\end{figure}

Selecting an optimal block size is critical for maximizing performance in the GPU-based QR algorithm on a given hardware platform. 
Note, although the triangular factor from \texttt{geqrf} can be reused in \texttt{orgqr}, the block size must remain consistent.
In practice, the optimal block size for \texttt{geqrf} is smaller than that for \texttt{orgqr}, which limits the performance of \texttt{orgqr}. 
Therefore, we recompute $\bm{T}^{-1}$ in \texttt{orgqr} routine.
Fig.~\ref{fig:mi210-v100-dgeqrf-dorgqr-nb-time} illustrates the performance tuning of our proposed \texttt{geqrf} and \texttt{orgqr} methods, evaluating various block sizes ($b$) for a fixed matrix size of $m$=20000 on MI210 and V100, with optimal elapsed times highlighted by larger symbols.
Furthermore, the results indicate that \texttt{geqrf} performs better on V100 than on MI210, attributed to V100’s superior BLAS2 performance, as evidenced by Fig.~\ref{fig:mi210-v100-dgemv-flops}, which forms part of its computational workload.
Conversely, \texttt{orgqr} exhibits higher performance on MI210, capitalizing on its enhanced BLAS3 performance, as shown in Fig.~\ref{fig:mi210-v100-dgemm-flops}, and our optimizations ensuring \texttt{orgqr} relies exclusively on 
BLAS3 operations.

\begin{figure}[ht]
    \centering
    \includegraphics[width=0.9\linewidth]{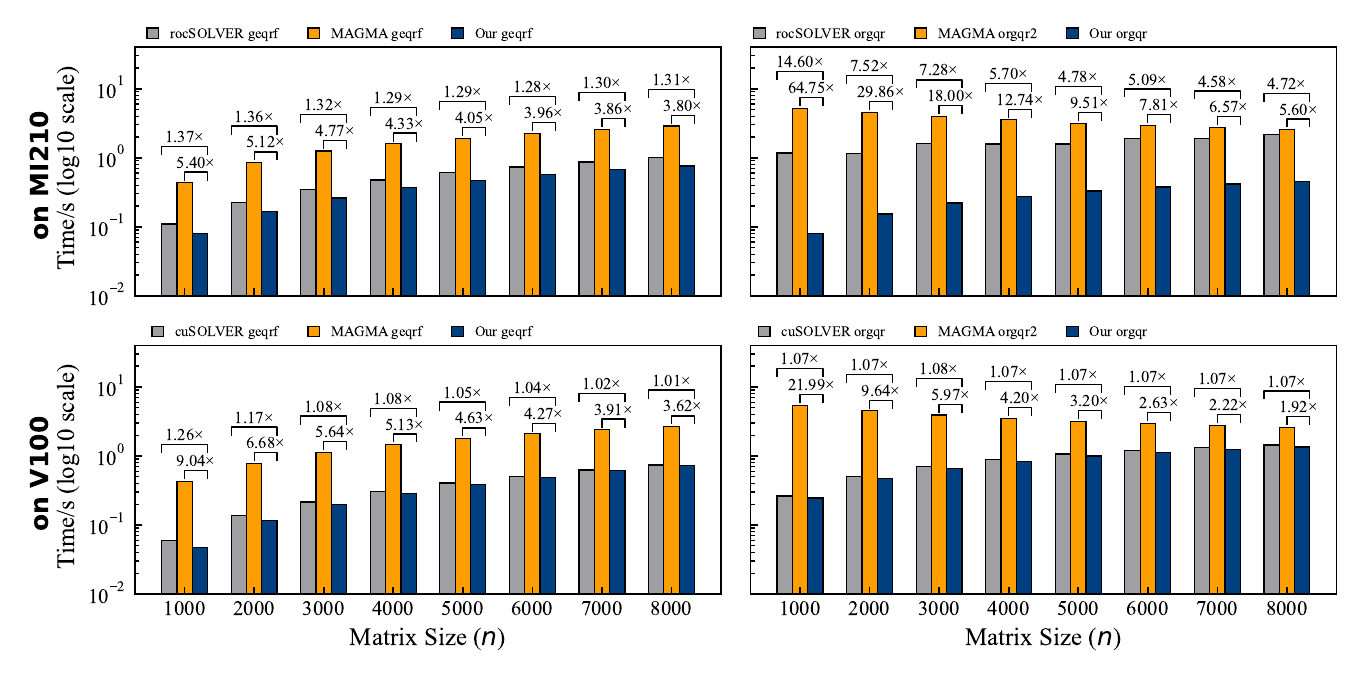}
    \caption{
    Performance comparison of \texttt{geqrf} and \texttt{orgqr}: rocSOLVER/cuSOLVER, MAGMA, and our proposed method for $m$=20000.
    }
    \label{fig:mi210-v100-dgeqrf-dorgqr-roc-cu-ma-our-m2w-time}
\end{figure}

Fig.~\ref{fig:mi210-v100-dgeqrf-dorgqr-roc-cu-ma-our-m2w-time} presents a performance comparison of \texttt{geqrf} and \texttt{orgqr} on MI210 and V100 GPUs for $m$=20000, evaluated across rocSOLVER/cuSOLVER, MAGMA, and our proposed method. The speedups achieved by our method relative to rocSOLVER/cuSOLVER and MAGMA are annotated above the bars.
As shown in Fig.~\ref{fig:mi210-v100-dgeqrf-dorgqr-roc-cu-ma-our-m2w-time}, our method consistently outperforms both rocSOLVER/cuSOLVER and MAGMA for \texttt{geqrf} and \texttt{orgqr} across all tested matrix sizes. 
Moreover, the speedup over MAGMA decreases as $n$ increases, indicating that our method is more suitable for taller-and-skinnier matrices.
It is worth noting that in MAGMA’s \texttt{magma\_dorgqr\_gpu}, the trailing part of matrix $\bm{Q}$--of size ($n$\%$b$+($m$-$n$))$^2$-- transferred from the GPU to the CPU for computation and then sent back, which incurs significant overhead when $m \gg n$. 
Instead, MAGMA uses \texttt{magma\_dgeqrf} and \texttt{magma\_dorgqr2} with the input matrix stored on CPU in the \texttt{magma\_dgesdd} routine.

\subsubsection{\textbf{Back Transformations on GPU}}
The \texttt{ormqr} and \texttt{ormlq} routines have accelerated versions  
available in MAGMA, where the trailing matrix update 
(\texttt{larfb})
is 
performed on GPU.
However, the 
generation of triangular factors (\texttt{larft}) is carried 
out on CPU,
necessitating CPU-GPU data transfers.
In our method, both \texttt{larft} and \texttt{larfb} 
are 
executed entirely on GPU, eliminating CPU-GPU data transfers.
Additionally, the optimization techniques, as described above, applied to \texttt{larft} and \texttt{larfb} can be extended to \texttt{ormqr} and \texttt{ormlq}, ensuring that all computations are performed using compute-bound BLAS3 operations, thereby maximizing GPU performance.

\begin{figure}[ht]
    \centering
    \includegraphics[width=0.9\linewidth]{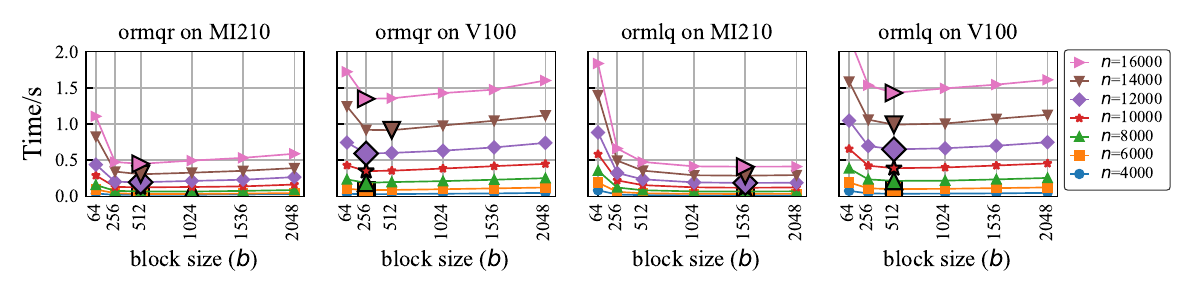}
    \caption{Tuning of our proposed \texttt{ormqr} and \texttt{ormlq} with varying block size ($b$).}
    \label{fig:mi210-v100-dormqr-dormlq-nb-time}
\end{figure}
\begin{figure}[ht]
  \centering
    \includegraphics[width=0.9\linewidth]{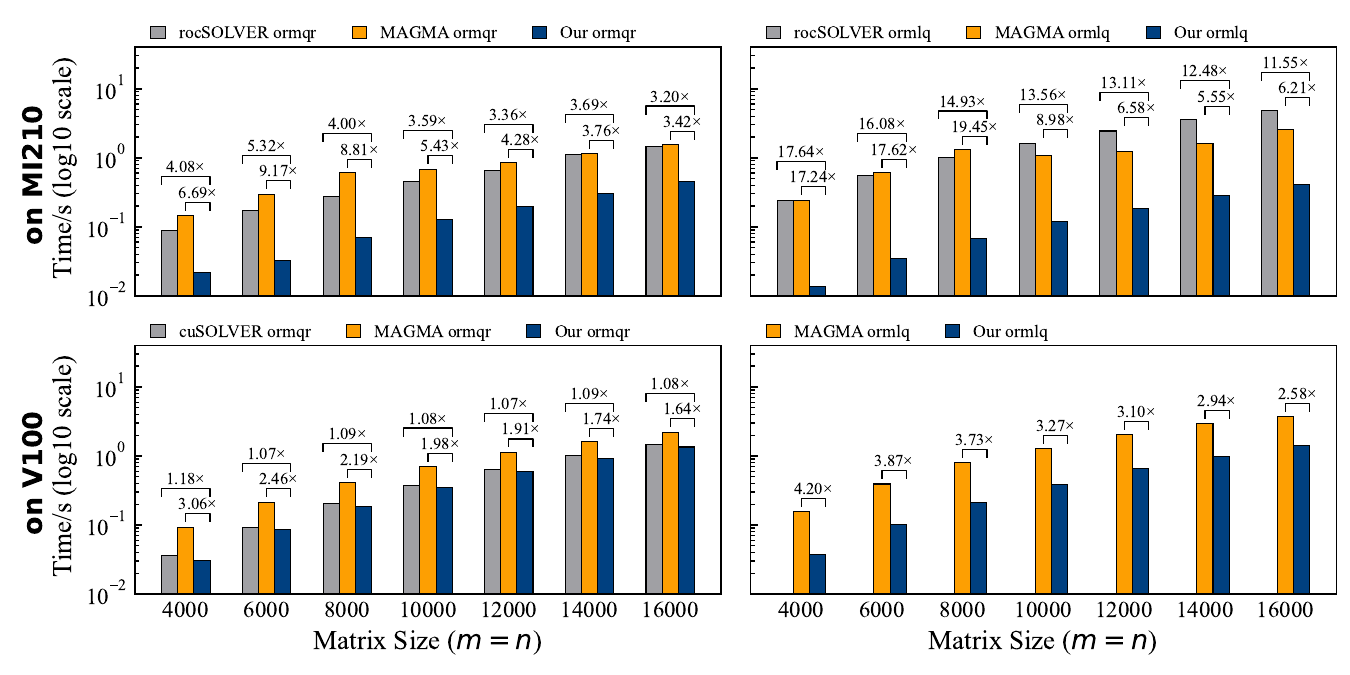}
    \caption{
    Performance comparison of \texttt{ormqr} and \texttt{ormlq}: rocSOLVER/cuSOLVER, MAGMA, and our proposed method.
    }
    \label{fig:mi210_dormbr_roc_ma_our}
\end{figure}
Fig.\ref{fig:mi210-v100-dormqr-dormlq-nb-time} shows the block size ($b$) tuning results for our \texttt{ormqr} and \texttt{ormlq} implementations, where larger markers indicate the optimal block size.
Fig.~\ref{fig:mi210_dormbr_roc_ma_our} compares the performance of \texttt{ormqr} and \texttt{ormlq} routines on MI210 and V100 across rocSOLVER/cuSOLVER, MAGMA, and our proposed methods for square matrices. 
Speedups over rocSOLVER/cuSOLVER and MAGMA are annotated above bars.
Notably, cuSOLVER does not provide an interface for \texttt{ormlq}. 
As shown, our optimized method consistently outperforms both rocSOLVER/cuSOLVER and MAGMA across all tested sizes on both GPUs.
The performance gains on MI210 are significantly higher than those on V100, which can be attributed to our methods’ exclusive reliance on BLAS3 operations, with MI210 offering superior BLAS3 performance.

\section{Combined experiment}\label{sec:svd-combined-result}

\subsection{Accuracy of SVD}
\begin{figure}[ht]
    {\includegraphics[width=0.8\linewidth]{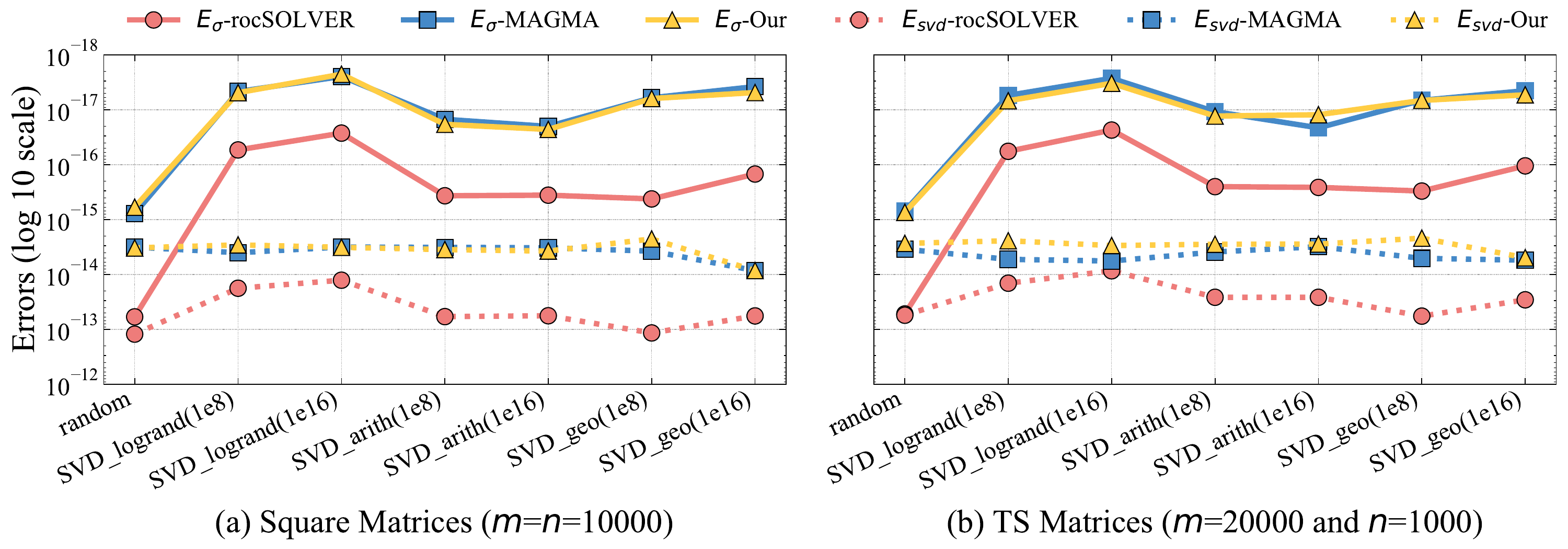}}
    \caption{Accuracy comparison of $E_{\sigma}$ (solid lines) and $E_{svd}$ (dashed lines) on MI210: rocSOLVER,  MAGMA and our SVD method.} \label{fig:mi210_svd_error_square_ts}
\end{figure}

The singular value error can be defined as:
\begin{displaymath}
    E_{\sigma} = \frac{\|\bm{\Sigma_{1}} -  \bm{\Sigma}_{2} \|_F}{n},
\end{displaymath}
where $\Sigma_{1}$ represents the reference singular values computed by LAPACK and the $\Sigma_{2}$ denotes the singular values obtained from rocSOLVER, MAGMA or our proposed method. If the SVD performed with singular vectors generated,
\begin{displaymath}
    E_{svd} =  \frac{\|\bm{A} - \bm{U} \times \bm{\Sigma} \times \bm{V}^\mathrm{T}\|_F}{\|\bm{A}\|_F}.
\end{displaymath}
Fig.~\ref{fig:mi210_svd_error_square_ts} presents the errors $E_{\sigma}$ and $E_{\text{svd}}$ for various matrix types with different condition numbers, including square ($m$=$n$=10000) and TS ($m$=20000, $n$=1000) matrices generated using the \texttt{magma\_generate\_matrix} function.
The results indicate that the accuracy of MAGMA and our proposed SVD method surpasses that of rocSOLVER, with our method achieving accuracy comparable to MAGMA. Overall, our SVD method exhibits robust stability.


\subsection{Profile of SVD Phases}

\begin{figure}[ht]
    \centering
    \begin{subfigure}[t]{0.85\textwidth}
    \centering
    {\includegraphics[width=\textwidth]{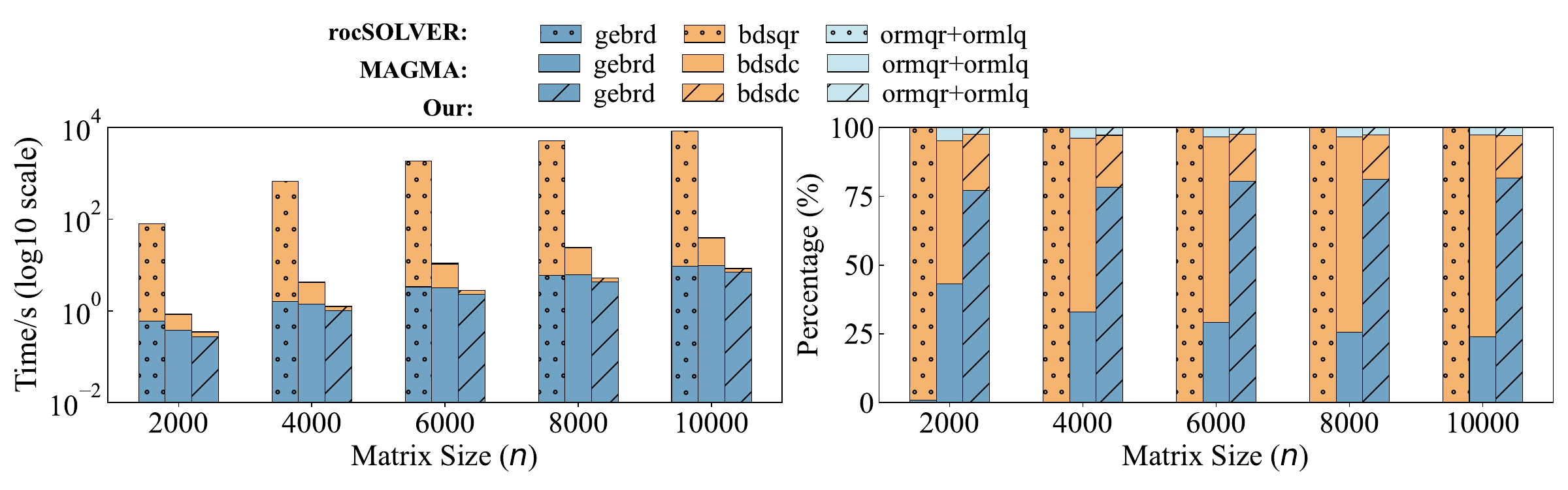}}
    \caption{Square Matrices}\label{mi210-svd-our-ma-roc-square-part}
    \end{subfigure}
    \begin{subfigure}[t]{0.85\textwidth}
    \centering
    \includegraphics[width=\linewidth]{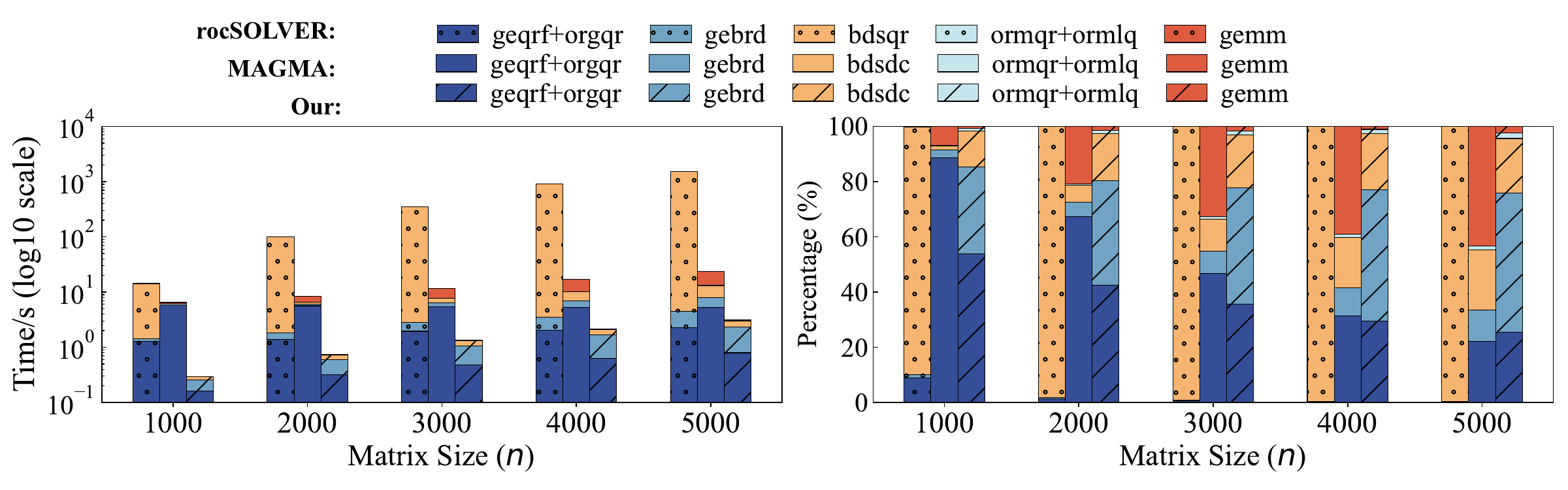}
    \caption{TS Matrices ($m$=20000)}\label{fig:mi210-svd-our-ma-roc-m20k-part}
    \end{subfigure}
    \caption{
    Profiling SVD Phases: Our vs. MAGMA vs. rocSOLVER on MI210.
    }    \label{fig:mi210-svd-our-ma-roc-part}
\end{figure}

To analyze the distribution of computation time across phases, we profile the SVD computation times for varying matrix sizes on MI210, comparing the performance of rocSOLVER
, MAGMA 
, and our proposed SVD method.
Fig.~\ref{fig:mi210-svd-our-ma-roc-part} illustrates
the computational time distribution for square matrices and TS matrices ($m$=20000). 
For square matrices, as shown in Fig.~\ref{mi210-svd-our-ma-roc-square-part},
the MAGMA implementation is dominated by the \texttt{gebrd} and \texttt{bdcdc} phases, accounting for 43.2\%$\sim$23.99\% and 51.8\%$\sim$73.3\% of the runtime, respectively, as matrix size increases, while \texttt{ormqr}+\texttt{ormlq} remains negligible at 4.9\%$\sim$2.7\%.
These results highlight the necessity of optimizing both the \texttt{gebrd} and \texttt{bdcdc} phases.
Our method achieves substantial speedups over MAGMA by optimizing all SVD phases, particularly \texttt{bdcdc}. 
Consequently, the contribution of the \texttt{bdcdc} phase is significantly reduced to 20.5\%$\sim$15.4\%, while \texttt{gebrd} becomes more dominant, increasing to 77.0\%$\sim$81.8\% as the matrix size grows. The \texttt{ormqr}+\texttt{ormlq} phase remains minor at 2.5\%$\sim$2.8\%.

For TS matrices ($m$=20000), as shown in Fig.~\ref{fig:mi210-svd-our-ma-roc-m20k-part}, 
the MAGMA method is initially dominated by the \texttt{geqrf}+\texttt{orgqr} phases, whose contribution decreases from 88.6\% to 22.2\% as n increases, while the \texttt{bdcdc} and final \texttt{gemm} phases increase from 1.4\% to 21.6\% and 6.9\% to 43.3\%, respectively.
The \texttt{gebrd} phase, though less pronounced, grows from 2.8\% to 11.4\% with increasing $n$, and the \texttt{ormqr}+\texttt{ormlq} phase remains minimal.
Our method accelerates all phases, particularly \texttt{geqrf}+\texttt{orgqr}, achieving greater speedups for taller and thinner matrices. 
In our approach, the \texttt{geqrf}+\texttt{orgqr} phases prevail for small $n$, decreasing from 53.9\% to 25.6\% as $n$ grows,
while the \texttt{gebrd} phase becomes dominant, rising from 31.4\% to 50.2\%.
The \texttt{bdcdc} phase remains a minor contributor, ranging from 13.0\% to 19.8\%, while the \texttt{ormqr}+\texttt{ormlq} and final \texttt{gemm} phases have minimal impact.


For both square and TS matrices, the time distribution across phases in rocSOLVER, as depicted in Fig.~\ref{fig:mi210-svd-our-ma-roc-part}, reveals that the \texttt{bdcqr} phase dominates execution time, underscoring it as the primary bottleneck and the key factor driving our method’s speedup over rocSOLVER.

\subsection{End-to-End SVD performance}
\begin{figure}[ht]
    \centering
    \includegraphics[width=0.8\linewidth]{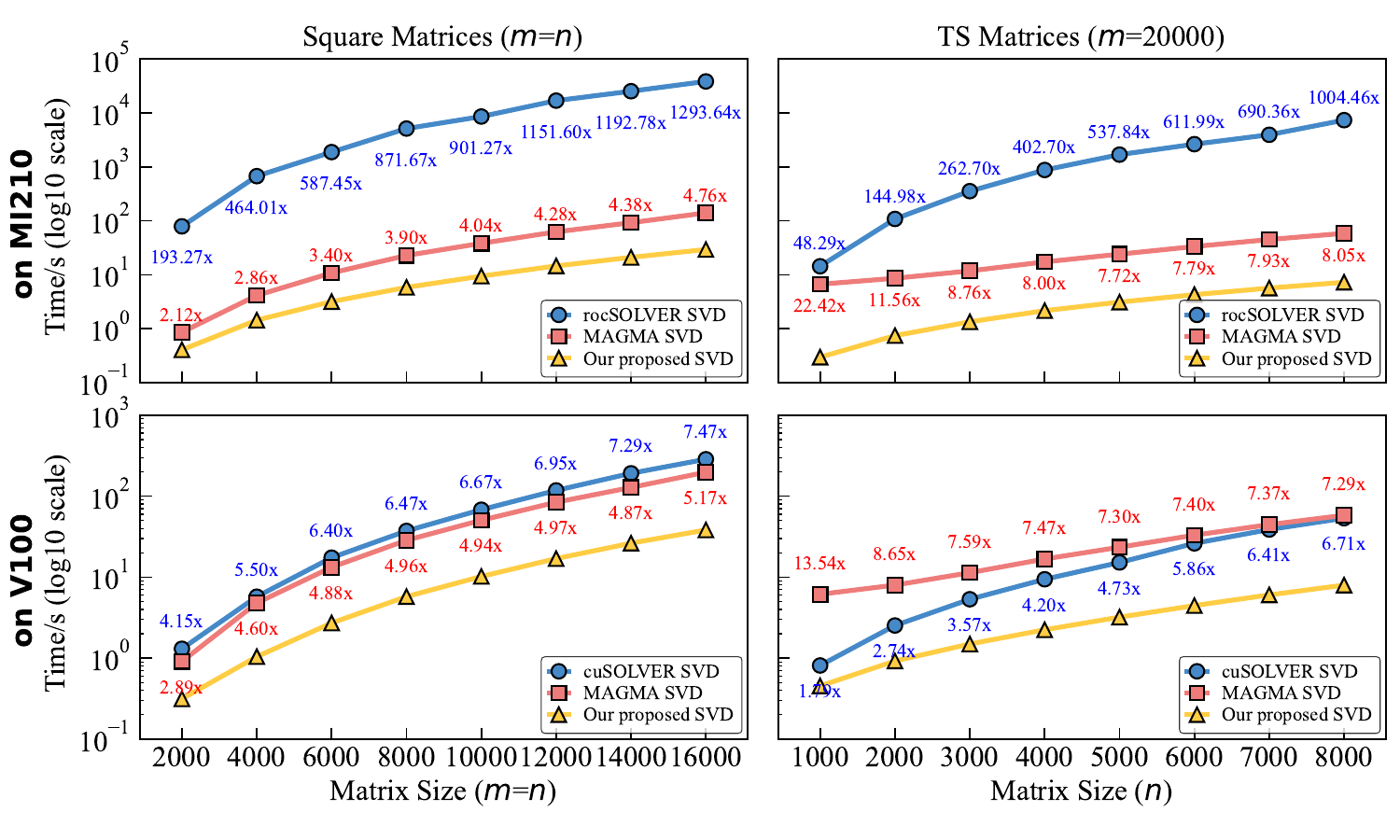}
    \caption{SVD Performance comparison: rocSOLVER/cuSOLVER, MAGMA, and our method.
    }
    \label{fig:mi210-v100-svd-roc-cu-ma-our-square-ts-20k}
\end{figure}

Fig.~\ref{fig:mi210-v100-svd-roc-cu-ma-our-square-ts-20k} compares the SVD performance between rocSOLVER/cuSOLVER, MAGMA and our proposed method for square and TS matrices ($m$=20000), with speedups indicated by numbers along the blue and red lines, respectively.
To achieve optimal performance, each phase employs the optimal block size identified in Section \ref{sec:svd-overall}. 
Our method consistently outperforms rocSOLVER/cuSOLVER across all evaluated matrix sizes, with speedups that increase with matrix dimensions, reaching up to 1293.64x and 7.47x, respectively. 
This is attributed to the enhanced efficiency of our GPU-based \texttt{bdcdc} approach compared to the \texttt{bdcqr} method in rocSOLVER/cuSOLVER.
For square matrices, the speedup over MAGMA increases with matrix size, achieving up to 4.76x on MI210 and 5.17x on V100.
For TS matrices, the speedup over MAGMA becomes more significant as $n$ decreases, highlighting the efficiency of our approach for taller-and-skinnier matrices.
Additionally, 
both our method and MAGMA perform better on MI210 than on V100.

\begin{figure}[ht]
    \centering
    \includegraphics[width=0.9\linewidth]{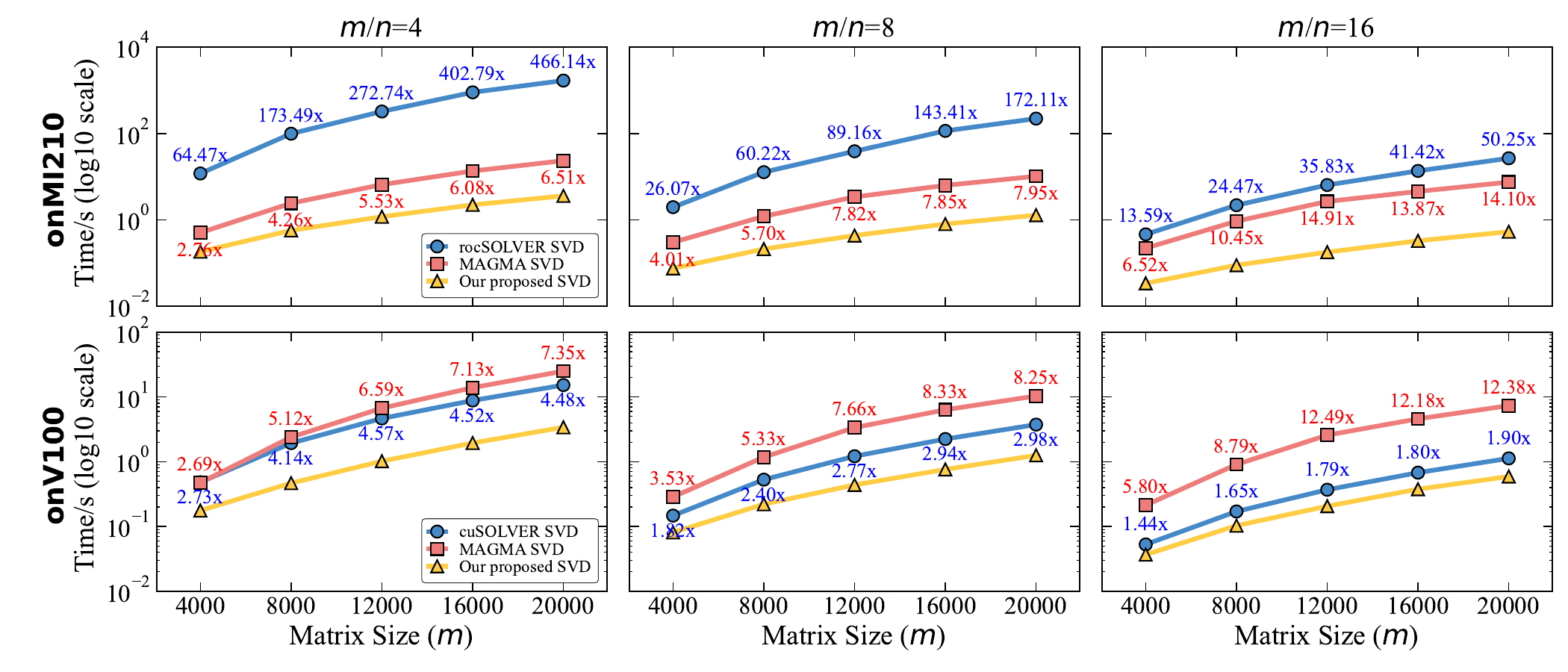}
    \caption{SVD Performance comparison for varying $m/n$ ratios: rocSOLVER/cuSOLVER, MAGMA, and our method.}
    \label{fig:mi210-v100-svd-roc-ma-our-time-mn-ratios}
\end{figure}
Fig.~\ref{fig:mi210-v100-svd-roc-ma-our-time-mn-ratios} compares the performance of our SVD method with rocSOLVER/cuSOLVER and MAGMA across different $m/n$ ratios (4, 8, and 16) on MI210 and V100, respectively. 
The Fig.~\ref{fig:mi210-v100-svd-roc-ma-our-time-mn-ratios} demonstrates that, for a fixed $m/n$ ratio, the speedup increases with matrix size ($m$), indicating greater acceleration for larger matrices. 
Similarly, when $m$ is held constant and the $m/n$ ratio increases, the speedup of our SVD method relative to MAGMA also increases, suggesting that taller and skinnier matrices benefit from more significant acceleration.
In contrast, as the $m/n$ ratio decreases, resulting in shorter and wider matrices, the speedup relative to rocSOLVER/cuSOLVER increases. This occurs because wider matrices lead to a higher proportion of computation time being dominated by \texttt{bdcqr} in rocSOLVER/cuSOLVER, where our GPU-based \texttt{bdcdc} approach provides a substantial performance advantage over \texttt{bdcqr}.

\section{Conclusion}


This paper presents a significant advancement in SVD through a GPU-centered algorithm designed to overcome the limitations of traditional approaches, such as slow panel factorization and frequent CPU-GPU data transfers in heterogeneous systems.
We reformulate the algorithm and data layout for key SVD stages—bidiagonalization, QR factorization, and singular vector back-transformations—to perform all panel-level computations and trailing matrix updates exclusively on GPU, eliminating CPU-GPU data transfers.
Additionally, we integrate related computations to optimize BLAS utilization, significantly increasing arithmetic intensity and fully leveraging GPU computational capabilities.
Furthermore, We propose a novel GPU-based bidiagonal divide-and-conquer (BDC) method that further enhances performance by restructuring the workflow to eliminate matrix-level data transfers and enable asynchronous CPU-GPU execution.
Extensive experiments on AMD MI210 and NVIDIA V100 GPUs demonstrate speedups of up to 1293.64x and 7.47x compared to rocSOLVER and cuSOLVER, respectively, and up to 14.10x and 12.38x relative to MAGMA, while preserving high numerical accuracy.
These results highlight the potential of our algorithm to establish a new benchmark for GPU-based SVD computations.

\begin{acks}
    This work is supported in part by the National Key R\&D Program of China (No. 2021YFB0300203), the National Natural Science Foundation of China (Nos. 12471348 and 12131005), and the Basic Research Project of Institute of Software, Chinese Academy of Sciences (No. ISCAS-PYFX-202302).
\end{acks}

\bibliographystyle{ACM-Reference-Format}
\bibliography{sample-base}

\appendix

\end{document}